\def \systemnameRaw {Talek}
\def \systemname {\systemnameRaw\xspace}
\def \name {\systemname}

\documentclass[sigconf]{acmart}

\copyrightyear{2020}
\acmYear{2020}
\setcopyright{rightsretained}
\acmConference[ACSAC 2020]{Annual Computer Security Applications Conference}{December 7--11, 2020}{Austin, USA}
\acmBooktitle{Annual Computer Security Applications Conference (ACSAC 2020), December 7--11, 2020, Austin, USA}
\acmDOI{10.1145/3427228.3427231}
\acmISBN{978-1-4503-8858-0/20/12}

\keywords{privacy, anonymity, messaging} 

\ccsdesc[500]{Security and privacy~Pseudonymity, anonymity and untraceability}
\ccsdesc[500]{Security and privacy~Privacy-preserving protocols}

\usepackage[small,compact]{titlesec}
\titlespacing*{\section}{2pt}{4pt}{1pt}
\titlespacing*{\subsection}{2pt}{2pt}{1pt}
\usepackage{etoolbox}
\makeatletter
\patchcmd{\ttlh@hang}{\parindent\z@}{\parindent\z@\leavevmode}{}{}
\patchcmd{\ttlh@hang}{\noindent}{}{}{}
\makeatother

\usepackage{array}
\usepackage{fancyhdr}
\usepackage{booktabs}
\usepackage{amsmath}
\usepackage{amsfonts}
\usepackage{calrsfs}
\usepackage{pifont}
\usepackage{verbatim}
\usepackage{url}
\usepackage{hyperref}
\hypersetup{
    colorlinks=false,
    pdfborder={0 0 0},
        }
\usepackage{graphicx}
\usepackage{caption}
\usepackage{subcaption}
\usepackage{sidecap}
\usepackage{float}
\usepackage[section]{placeins}
\usepackage{enumitem}
\usepackage{color}
\usepackage{listings}
\usepackage{algorithm}
\usepackage{algpseudocode}
\usepackage{multirow}
\usepackage{xspace}
\usepackage{microtype}

\lstdefinelanguage{javascript}{
  keywords={typeof, new, true, false, catch, function, return, null, catch, switch, var, if, in, while, do, else, case, break},
  keywordstyle=\color{blue}\bfseries,
  ndkeywords={class, export, boolean, throw, implements, import, this},
  ndkeywordstyle=\color{black}\bfseries,
  identifierstyle=\color{black},
  sensitive=false,
  comment=[l]{//},
  morecomment=[s]{/*}{*/},
  commentstyle=\color{purple}\ttfamily,
  stringstyle=\color{red}\ttfamily,
  morestring=[b]',
  morestring=[b]"
}

\newcommand{\squishlist}{\begin{itemize}[itemsep=0pt,parsep=0pt,topsep=0pt,partopsep=0pt,leftmargin=0em,labelwidth=1em,labelsep=0.5em]}
\newcommand{\squishlistend}{\end{itemize}}
\newcommand{\squishend}{\end{itemize}}
\newcommand{\squishenum}{\begin{enumerate}[itemsep=0.5pt,parsep=0pt,topsep=0pt,partopsep=0pt,leftmargin=1.5em,labelwidth=1em,labelsep=0.5em]{}}
\newcommand{\squishenumend}{\end{enumerate}}
\newcommand{\squishenuma}{\begin{enumerate}[label=(\alph*),itemsep=0.5pt,parsep=0pt,topsep=0pt,partopsep=0pt,leftmargin=1.5em,labelwidth=1em,labelsep=0.5em]{}}
\newcommand{\squishenumenda}{\end{enumerate}}
\newcommand{\todo}[1]{[[[{\bf{TODO: #1}}]]]}
\newcommand\myurl[1]{\url{#1}}

\newcommand{\captionfonts}{\normalsize}
\makeatletter  \long\def\@makecaption#1#2{  \vskip 0.1in
  \sbox\@tempboxa{{\captionfonts #1: #2}}  \ifdim \wd\@tempboxa >\hsize
    {\captionfonts #1: #2\par}
  \else
    \hbox to\hsize{\hfil\box\@tempboxa\hfil}  \fi
  \vskip 0in}
\makeatother

\newif\ifsubmit    \submittrue
\ifsubmit
  \renewcommand{\todo}[1]{}
\else
  \renewcommand{\todo}[1]{\ding{110}\ding{43}\textcolor{red}{#1}}
\fi

\begin{document}

\title{\systemname{}: Private Group Messaging with Hidden Access Patterns}

\author{Raymond Cheng}
\affiliation{University of Washington}
\email{ryscheng@cs.washington.edu}
\author{William Scott}
\affiliation{University of Washington}
\email{wrs@cs.washington.edu}
\author{Elisaweta Masserova}
\affiliation{Carnegie Mellon University}
\email{elisawem@andrew.cmu.edu}
\author{Irene Zhang}
\affiliation{Microsoft Research}
\email{irene.zhang@microsoft.com}
\author{Vipul Goyal}
\affiliation{Carnegie Mellon University}
\email{vipul@cmu.edu}
\author{Thomas Anderson}
\affiliation{University of Washington}
\email{tom@cs.washington.edu}
\author{Arvind Krishnamurthy}
\affiliation{University of Washington}
\email{arvind@cs.washington.edu}
\author{Bryan Parno}
\affiliation{Carnegie Mellon University}
\email{parno@cmu.edu}

\renewcommand{\shortauthors}{R. Cheng, W. Scott, E. Masserova, I. Zhang, V. Goyal, T. Anderson, A. Krishnamurthy, and B. Parno}

\begin{abstract}

\systemname{} is a private group messaging system that sends messages through potentially untrustworthy servers, while hiding both data content and the communication patterns among
its users. \systemname{} explores a new point in the design space of private messaging; it guarantees access sequence indistinguishability, which is among the strongest guarantees in the space, while assuming an anytrust threat model, which is only slightly weaker than the strongest threat model currently found in related work. Our results suggest that this is a pragmatic point in the design space, since it supports strong privacy \emph{and} good performance: we demonstrate a 3-server \systemname{} cluster
that achieves throughput of 9,433 messages/second for 32,000 active users with
1.7-second end-to-end latency. 
To achieve its security goals without coordination between clients, \systemname{} relies on
information-theoretic private information retrieval.
 To achieve good performance and minimize server-side storage, \systemname{} introduces new 
techniques and optimizations that may be of independent interest, 
e.g., a novel use of blocked cuckoo hashing and 
support for private notifications. 
The latter provide a private, efficient mechanism for users to learn, without polling,
which logs have new messages.

\end{abstract}

 \maketitle
\vspace{-1.0em}
\section{Introduction}\label{sec:introduction}
Messaging applications depend on cloud servers to exchange data,
giving server operators full visibility into the communication patterns between users.
Even if the communication contents are encrypted,
network metadata can be used to infer
which users share messages, when traffic is sent, where data is sent, and how much is transferred.
This can allow the servers and/or network providers
to infer the contents of the communication~\cite{islam2012access}.
When remote hacking, insider threats, and government requests are common,
protecting the privacy of communications requires that we guarantee security
against a stronger threat model.
For some users, e.g., journalists and activists,
protecting metadata is critical to their job function and
safety~\cite{mcgregor2015investigating,mcgregor2016individual}.

As we describe in \S\ref{sec:related}, a wide variety of systems explore ways of protecting the privacy of such metadata. 
We can classify this prior work into two groups based on the privacy \emph{guarantees} offered
and the \emph{threat model} each system defends against.
The first group of work~\cite{alexopoulos2017mcmix,angel2018,angel2016unobservable,goldreich1987towards,goldreich1996software,ostrovsky1990efficient,borisov2015dp5,gupta2016scalable} offers strong security guarantees against very strong threat models 
(e.g., assuming that only the clients themselves are trusted).
Unfortunately, this typically imposes prohibitive computational or network costs.
The second group~\cite{corrigan2015riposte,corrigan2010dissent,wolinsky2013proactively,kwon2016riffle,
wolinsky2012dissent,kwon2017atom,cwtch,pond-tor,ricochet-tor,sassaman2005pynchon,van2015vuvuzela,tyagi2017stadium,kwon2019xrd,piotrowska2017loopix} 
offers weaker security guarantees (such as k-anonymity~\cite{sweeney2002k}, plausible deniability~\cite{hong2004architecture}
or differential privacy~\cite{dwork2006differential,dwork2008differential}) 
and often much weaker threat models too; e.g., a fraction of the servers must be honest.
However, in exchange for weakening the guarantees and threat model,
these systems often achieve impressive performance results.

In this work, we explore an intriguing middle ground: we define \emph{access sequence indistinguishability}, 
a notion similar to (but slightly stronger than) the security guarantees from systems in the first group, and combine it with an ``anytrust'' threat model~\cite{anytrust},
which is slightly weaker than the threat models of the first group, but stronger than those of the second  group. Intuitively, the anytrust threat model assumes that different organisations, e.g., Mozilla, EFF, and WikiLeaks, each provide servers and at least one (unknown) organisation can be trusted.
The hope is that we can promise the strong guarantees of the first group
and the strong performance of the second, 
while defending against a strong threat model.

To explore this design point, we construct \name, a private communication system
targeting small groups of trusted users communicating amongst themselves
(e.g., friends chatting via IRC or text messaging). 
To support such applications,
\name offers the abstraction of a private log with a single writer and multiple readers.
Clients store and retrieve asynchronous messages on untrusted servers without revealing any
communication metadata.
If the group of friends and at least one server are uncompromised,
\systemname{} prevents an adversary from learning anything about their communication patterns.
Combined with standard message encryption, 
\name conceals both the \emph{contents} and \emph{metadata} of clients' application usage without sacrificing
cloud reliability and availability.

Similar to prior systems,
to hide communication patterns, 
\name clients issue fix-sized, random-looking network requests 
at a rate which is independent of application-level requests.
Hence, application-level requests must occasionally be delayed,
and ``dummy'' network requests must be issued when no application
requests are ready.
As with any privacy system, careful application-specific tuning
is necessary to trade off between the amount of cover traffic sent
and the latency of real application requests.

Unlike prior systems with \name's strong guarantees, to achieve good performance, \name leverages the anytrust
threat model, which allows us to use \emph{information theoretic} private
information retrieval (IT-PIR)~\cite{chor1998private,devet2012optimally,goldberg2007improving}.
IT-PIR requires an anytrust assumption,
but in exchange it avoids the use of heavyweight crypto operations required for other flavors of PIR~\cite{kushilevitz1997replication}.
Abstractly, PIR allows a client to retrieve the $i$-th record from 
a ``database'' of $n$ items held collectively by $l$ servers,
without the servers learning which record was retrieved.
However, PIR alone (of any flavor) is not enough to support efficient group messaging.
In particular, it does not explain how messages are privately written to the servers,
how readers find messages sent to them,
nor how to structure the PIR database to facilitate maximum efficiency.

In \name, within a group of clients reading and writing to a message log, a shared secret
determines a pseudorandom, deterministic sequence of database locations for messages.
Any user with the log secret (a capability) can follow the pseudorandom sequence,
reading new messages independently.
One of \name's key innovations is to show how
these messages can be stored in a \emph{blocked cuckoo hash table}~\cite{dietzfelbinger2007balanced} 
to provide efficient time and space usage in the context of IT-PIR.
We further optimize read performance via a novel technique - \emph{private notifications}. 
Private notifications allow users who subscribe to multiple logs
to learn, in a privacy-preserving fashion, which of their logs have new messages.
Hence, they can internally prioritize reading from those logs
and avoid inefficient polling of logs that have no new messages.

Like any privacy system,
\name has several limitations.
First, the current implementation of \name does not guarantee liveness;
any user in a trusted group can block writes to that group's log. 
Similarly, a faulty server can impact availability.
Clients can detect but not attribute such faults.
Hence, service providers should be chosen with reputations for high availability. An alternative is to use robust versions of PIR~\cite{devet2012optimally,goldberg2007improving} at the cost of higher overhead. We defer investigating this option to future work.
Second, because clients connect directly to \systemname{} servers, as with other practical systems~\cite{van2015vuvuzela,angel2016unobservable,kwon2016riffle,angel2018,kwon2017atom,tyagi2017stadium,lazar2018karaoke,kwon2019xrd}, we do not hide when users are online.
Orthogonal systems (e.g., Tor~\cite{dingledine2004tor}) may help here.
Third, as discussed earlier, we expect \name to be used for communication among small groups of trusted users.
If a log secret is shared with the adversary, writer anonymity for that log is compromised,
but readers' anonymity and writer anonymity for other logs are preserved.
Applications that require broadcasts to many untrusted users (e.g., a public blog)
are better served by anonymous broadcast systems~\cite{corrigan2015riposte,corrigan2010dissent,wolinsky2013proactively,wolinsky2012dissent}.

We have implemented two versions of \systemname{}, one entirely in Go 
and one that offloads PIR operations to a GPU;
our code is publicly available.
We evaluated the system on a 3-server deployment using Amazon EC2. 
To provide a realistic group messaging workload,
we replay the Ubuntu IRC message logs from 2016~\cite{ubuntu-irc}.
Overall, we find that this design point is surprisingly practical.
Even with 32,000 clients actively reading and writing messages according to a
fixed schedule every second,
we show that clients use 148MB per day to achieve an average
end-to-end message latency of 1.7 seconds, measured from the time a sender enters
a message to the time the recipient sees it.
Under this workload, our server supports a peak throughput of 9,433 messages per second,
orders of magnitude better performance than systems with similar security goals.

In summary, we make the following contributions.
\squishenum{}
  \item Talek, a system that explores an important design point within private group messaging with strong guarantees, strong threat model, and high performance.
  \item A novel use of blocked cuckoo hashing for IT-PIR.
  \item Private notifications which privately encode the set of new messages, helping clients prioritize reads.
  \item Two open-source implementations of \name exploring the tradeoffs between CPU and GPU-based computation.
 
\squishenumend{}

 \section{Background: PIR}
\label{sec:background}
\label{sec:background:pir}

\systemname{} uses the privacy guarantees
of PIR in the context of a group messaging protocol.
PIR allows a single client to retrieve a block from a set of storage replicas
without revealing to any server the blocks of interest to the client.
There exist two major categories of PIR techniques,
computational PIR (C-PIR)~\cite{kushilevitz1997replication}
and information-theoretic PIR (IT-PIR)~\cite{chor1998private,goldberg2007improving,devet2012optimally}.
\systemname{} is compatible with both varieties.
Prior work~\cite{angel2016unobservable,angel2018} 
explored C-PIR, since it supports the strongest possible threat model: 
only the communication partners must be trusted.
However, their results indicate that C-PIR imposes significant
computation and network overheads, raising practicality concerns.
With \name, we focus on IT-PIR, which requires an anytrust threat model
but holds the promise of better performance.

To provide intuition for the performance and cost of IT-PIR,
we illustrate a standard protocol~\cite{chor1998private} with an example.
Let $l$ represent the number of servers, each storing a full copy of the database,
partitioned into equal sized blocks.
While IT-PIR generalizes to arbitrary numbers of servers and blocks,
our example contains $l=3$ servers and
$n=3$ blocks ($\{B_1, B_2, B_3\}$).

\squishenum{}
  \item     Suppose a client wants to read the second block, $\beta=2$, encoded by the bit vector, $q'=[0,1,0]$,
    which consists of zeros and a one in position $\beta$.
  \item     The client generates $l-1$ (e.g., 2) random $n$-bit request vectors, $q_1$ and $q_2$.
  \item     The client computes the last request vector as the XOR of the vectors from (1) and (2),
    $q_{l} = q' \oplus q_1 \oplus \ldots \oplus q_{l-1}$.
  \item     The client then sends $q_i$, to server $i$ for $1 \leq i \leq l$.
    Since request vectors are generated randomly, this reveals no information to any collection of $< l$ colluding servers.
  \item     Suppose $B_j$ represents the $j^{th}$ block of the database.
    Each server $i$ receives $q_i = [b_1, \ldots, b_n]$ and computes $R_i$, the XOR of all $B_j$ for which $b_j==1$
    and returns $R_i$ to the client.
  \item     The client restores the desired block, $B_{\beta}$, by taking the XOR of all $R_i$, i.e.,
    $B_{\beta} = R_1 \oplus \ldots \oplus R_l$ (since $q' = q_1 \oplus \ldots \oplus q_l$, and $q'$ is all zeroes except at index $\beta$).
\squishenumend{}

IT-PIR has desirable network properties: a client sends one request vector
and receives one block from each server. These requests and responses appear
random to the network and the servers, assuming at least one server is honest.
The size of a client request scales with total number of blocks,
and the client work scales with the number of servers.

For the servers,
IT-PIR can be computationally expensive
even though the individual operations (XOR) are cheap,
since the computational cost
for a read request scales at least linearly with the size and number of blocks in the system. 
Theoretical work suggests this limitation is inherent (\S\ref{sec:related}).
IT-PIR also requires consistent snapshots across servers,
with equal sized blocks in the data structure.

\section{Problem Definition \& Threat Model}
\label{sec:problem}
\label{sec:background:threat}

\begin{figure}
  \begin{center}
    \includegraphics[width=0.4\textwidth]{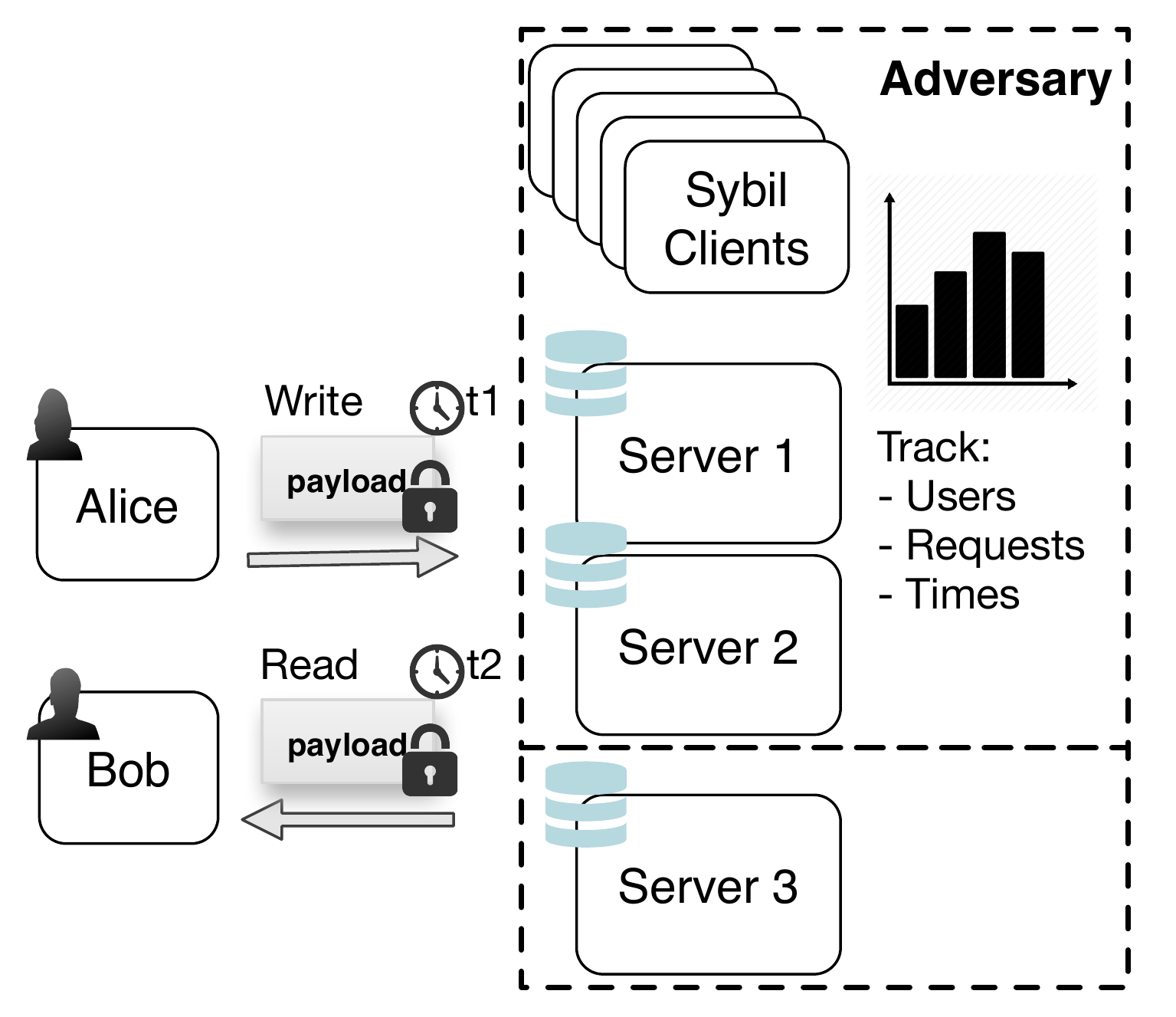}
  \end{center}
  \vspace{-1.0em}
  \caption{\footnotesize
    \textbf{\name's System and Threat Model.} 
    We assume the adversary can control all but one of $l$ servers in the system (here, $l=3$).
    Clients send network requests directly to the servers.
    Adversarial servers are free to record additional data, such as
    the source, type, parameters, timing, and size of all requests
    to link users who are likely to be communicating together.
  }
  \label{fig:networkmodel}
\end{figure}

Figure~\ref{fig:networkmodel} illustrates our problem setting:
various clients located across a wide-area network
communicate via a messaging service.
The service is provided by 
$l$ servers each controlled by a different administrative domain.

\label{sec:background:goal}

We wish to prevent the adversary from learning any
information about the communication patterns amongst honest users.
While not described in this paper, other end-to-end messaging guarantees~\cite{unger2015sok},
such as message integrity, authentication~\cite{bellare1996keying,rivest1978method},
forward secrecy~\cite{signal-privacy-policy},
and fork consistency~\cite{li2004sundr,mahajan2011depot},
can be provided by \systemname{} by including additional data in the message payload.
We focus on the privacy of access sequences,
since that forms the foundation for private messaging.

We provide a model of client and server interactions (\S~\ref{sec:syntax}),
and then formally introduce \emph{access sequence indistinguishability} (\S~\ref{sec:secgame}). We define it as a game played in rounds between an adversary and a challenger.  The adversary controls some number of dishonest clients and servers, the challenger controls some number of honest clients and randomly chooses a challenge bit which is fixed for the duration of the game. Informally, in each round, the adversary specifies two actions for each honest client, the challenger chooses one according to the challenge bit and adds it to the request queue of the corresponding client. Then, the challenger executes all requests that must be executed in this round as specified by the real-world protocol.
Intuitively, the notion of access sequence indistinguishability means that an adversary cannot
distinguish between an honest user's access patterns and a
random access pattern of arbitrary length.

Access sequence indistinguishability 
provides one of the strongest definitions of privacy available in private group messaging.
It is reminiscent of the definitions used in early works on 
oblivious RAM (ORAM)~\cite{goldreich1987towards,goldreich1996software,ostrovsky1990efficient},
but it supports multiple distinct readers and writers.
It is stronger than k-anonymity~\cite{sweeney2002k},
where the adversary can narrow the user's identity down to one of $k$ users 
(where for performance, $k$ is typically much smaller than the number of online users).
It is also stronger than plausible deniability~\cite{hong2004architecture},
where information leakage is allowed up to a certain confidence bound. 
It is different from unobservability~\cite{hevia2008indistinguishability}, in that we do not hide the maximum number of messages sent by the user, but we do account for the order of the messages in the access sequences. 
It is most similar to the notion of 
relationship unobservability under explicit retrieval (UO-ER)~\cite{angel2016unobservable,angel2018},
which is based on the notion of relationship unobservability \cite{pfitzmann2010terminology}.
We believe access sequence indistinguishability offers a cleaner definition, however,
and differs in at least two small, concrete ways:
\squishenum{}
  \item In UO-ER, at least one message is sent \emph{and} received in each round. In access sequence indistinguishability, the send and receive rates are decoupled. 
  \item In access sequence indistinguishability, the adversary is fully adaptive, i.e., the adversary can observe the network events and modify its strategy in each round. 
        In UO-ER, each time the adversary adapts its strategy,
        the state of the clients resets, which may reduce 
        the adversary's utility from acting adaptively.
\squishenumend{}

\subsection{Model}\label{sec:syntax}

Our setting consists of a finite set of servers and clients,
and we assume that all parties are stateful. 
The overall system consists of the following, possibly randomized and interactive, subroutines:

\vspace{0.03in}
\noindent {\it $\mathit{RealWrite}(\tau, seqNo, M) \rightarrow \{\omega^{0}, \ldots, \omega^{l-1}\}$:}
Clients use the RealWrite function to generate Write requests
sent to the $l$ servers. The RealWrite function takes as input
a log handle $\tau$ and a message $M$ with the sequence number $seqNo$ to publish to the log,
producing a set of $l$ Write requests, one per server. 

\vspace{0.03in}
\noindent {\it $\mathit{FakeWrite()} \rightarrow \{\omega^{0}, \ldots, \omega^{l-1}\}$:}
Clients use the FakeWrite function to generate a set of random ``dummy'' Write requests,
one per server.

\vspace{0.03in}
\noindent {\it $\mathit{RealRead}(\tau, seqNo) \rightarrow \{q^{0}, \ldots, q^{l-1}\}$:}
Clients use the RealRead function to generate Read request queries sent to the servers.
It takes as input a log handle $\tau$ and a sequence number $seqNo$ in the log,
producing a set of $l$ Read requests, one per server.

\vspace{0.03in}
\noindent {\it $\mathit{FakeRead()} \rightarrow \{q^{0}, \ldots, q^{l-1}\}$:}
Clients use FakeRead to generate a set of random Read requests, one per server.

\vspace{0.03in}
\noindent {\it $\mathit{GetUpdates()} \rightarrow V$:}
Clients use the GetUpdates function (modeled as an interactive subroutine) to retrieve the global interest vector, $V$, from the servers. The global interest vector encodes the set of (log handle, message sequence number)-pairs for all messages currently on the server.

\vspace{0.03in}
\noindent {\it $\mathit{ProcessWrite}(\omega^i) \rightarrow \bot$:}
Server $i$ uses the ProcessWrite function to process an incoming write request, $\omega^i$,
and update the server's internal state. 

\vspace{0.03in}
\noindent {\it $\mathit{ProcessRead}(q^i) \rightarrow R^i$:}
Server $i$ uses the ProcessRead function to process incoming read requests.
The function takes as input a read request, $q^i$, and outputs a reply $R^i$.

\vspace{0.03in}
\noindent {\it $\mathit{ProcessGetUpdates()} \rightarrow V^i$:}
Server $i$ uses ProcessGetUpdates to generate a global interest vector, $V^i$.

\subsection{Security Game Definition}
\label{sec:secgame}
We define access sequence indistinguishability using the following security game,
played between the adversary, $\mathcal{A}$, and a challenger, $\mathcal{C}$.
$\mathcal{A}$ is a probabilistic, polynomial-time adversary
who is in control of the network, $t$ out of $l$ servers (for Talek, $t=l-1$),
and a polynomial number of clients.
$\mathcal{A}$ can drop any message, send arbitrary messages from any of the adversarial clients to any server,
respond arbitrarily to requests, and modify any server-side state for adversarial servers.
The challenger, in turn, emulates (internally) the honest clients and servers.

Intuitively, the game allows the adversary to repeatedly, dynamically, and adaptively 
specify any two actions for each legitimate client to take.
The challenger has the client execute one of the two actions,
based on the random bit $b$ chosen for the game,
and the adversary can then choose new actions for the legitimate clients
and/or have its malicious clients and servers take actions.
This continues until the adversary produces a guess for the value of $b$.
If the adversary does no better than random chance,
this implies that the adversary cannot determine which users access the same logs, 
because the adversary could have chosen to have 
client actions with overlapping logs across users. 

In the definition, for simplicity, we assume the presence of authenticated
secure channels between each client-server pair (e.g., with TLS) and omit the
corresponding operations. 

\renewcommand{\squishlist}{\begin{itemize}[itemsep=0pt,parsep=0pt,topsep=0pt,partopsep=0pt,leftmargin=-1mm,labelwidth=1em,labelsep=0.5em]}

\squishenum
  \item $\mathcal{A}$ chooses a non-negative integer, $m$, and submits this number to the challenger,
    who spawns $m$ clients, $\mathcal{C}_0 \ldots \mathcal{C}_{m-1}$.
  \item The challenger flips a coin, $b \in \{0, 1\}$, uniformly at random,
    which is fixed for the duration of the game.
  \item For each of the challenger's clients, $\mathcal{C}_j$,
    $\mathcal{A}$ maintains two unique data access sequences, $seq^0_j$ and $seq^1_j$.
  \item The challenger maintains a table $T$ of log handles that have been created.  $T$ is initially empty.
  \item In each round, until $\mathcal{A}$ ends the game, one of the following four actions happens:
\squishenuma
  \item \textbf{$\mathcal{A}$ creates a log}
    \squishlist 
  \item $\mathcal{A}$ submits a create log request to the challenger. This request specifies the index of a client acting as a writer of the log, and the indices of the clients acting as readers of the log.
  \item $\mathcal{C}$ checks the client indices provided in the create log request, and if all indices correspond to the clients spawned by the challenger, $\mathcal{C}$ generates a log handle $\tau$. $\mathcal{C}$ saves $\tau$ in $T$ and returns the corresponding index $ind$ in $T$ to the adversary. 
    \squishend
  \item \textbf{$\mathcal{A}$ causes clients to $\mathit{GetUpdates}$}
    \squishlist
    \item $\mathcal{A}$ identifies a set of clients $I$.
    \item Clients in $I$ execute $\mathit{GetUpdates}$.  Challenger-controlled servers respond with $\mathit{ProcessGetUpdates}$. 
    \squishend
  \item \textbf{$\mathcal{A}$ extends the access sequences}
    \squishlist
    \item For all $m$ challenger clients, $\mathcal{A}$ chooses the $i$-th operation for both sequences:
      $\{seq^0_0[i] \ldots seq^0_{m-1}[i]\}$ and \\
      $\{seq^1_0[i], \ldots seq^1_{m-1}[i] \}$.
      $\mathcal{A}$ submits operations $seq^0_j[i]$ and $seq^1_j[i]$ to the respective client, $\mathcal{C}_j$.
      An operation can be $(ind, \mathit{RealWrite}(\cdot, seqNo, M))$, $(ind, \mathit{RealRead}(\cdot, seqNo))$,\\
      $\mathit{FakeWrite()}$, or $\mathit{FakeRead()}$, where $ind$ is the index of the log handle for the request.
    \item Each client, $\mathcal{C}_j$, receives one operation: $seq_j^b[i]$. If it is a $\mathit{FakeRead()}$ or $\mathit{FakeWrite()}$ operation,  $\mathcal{C}_j$ adds it to its read- or write queue, respectively. Otherwise, the challenger looks up the index $ind$ in table $T$. If it finds the index and corresponding log handle $\tau$, client $\mathcal{C}_j$ sets the log handle of $seq_j^b[i]$ to $\tau$ and adds this updated operation to the read queue if $seq_j^b[i]$ is a $\mathit{RealRead}$, and to the write queue if it is a $\mathit{RealWrite}$.
    \item If a write request must be issued this round (according to the write rate $w$), each challenger client $\mathcal{C}_j$ dequeues a request from its write queue and executes it. If the write queue is empty, $\mathcal{C}_j$ executes a $\mathit{FakeWrite}$.
    \item If a read request must be issued this round (according to the read rate $r$), each challenger client $\mathcal{C}_j$ dequeues a request from its read queue and executes it. If the read queue is empty, $\mathcal{C}_j$ executes a $\mathit{FakeRead}$.   
    \item Challenger-controlled servers use the $\mathit{ProcessRead}$ and \\
      $\mathit{ProcessWrite}$ routines to respond to the corresponding requests.
    \item Adversary-controlled clients can send arbitrary requests to any server.
      Adversary-controlled servers can modify their own state and respond arbitrarily.
    \item $\mathcal{A}$ observes the network events, $events^{b'}_j[i]$ sent from
      $\mathcal{C}$'s clients to all servers,
      including the outputs of $\mathit{RealWrite}$, $\mathit{FakeWrite}$, $\mathit{RealRead}$, and $\mathit{FakeRead}$.
  \squishend
  \item \textbf{$\mathcal{A}$ ends the game with its guess, $b'$, for $b$.}
  \squishenumenda
\squishenumend
\renewcommand{\squishlist}{\begin{itemize}[itemsep=0pt,parsep=0pt,topsep=0pt,partopsep=0pt,leftmargin=1em,labelwidth=1em,labelsep=0.5em]}

\vspace{-0.5em}
\begin{definition} (Access Sequence Indistinguishability)
  The system provides access sequence indistinguishability
  with security parameter $\lambda$ 
  if for any polynomial-time probabilistic adversary
  \begin{align*}
    |Pr(b=b') - 1/2| \leq negl(\lambda)
  \end{align*}
  in the security game, where $negl$ is a negligible function.
\end{definition}
\vspace{-0.5em}

In the game, when the adversary asks to create a log,
we restrict the members of the group to honest clients,
which corresponds directly to the trusted group assumption. 
We do not hide retrievals of the global interest
vector via $\mathit{GetUpdates}$, but we do hide the fact that a user of a
trusted group is executing a real read or write request.
In Talek, $t=l - 1$, since this corresponds to the anytrust threat model.

\paragraph{Availability}\label{sec:background:availability}
Threats to availability are out of scope for this paper.
During normal operation, all servers must be available and reachable by the clients.
Each server can deny service availability by refusing to respond 
or by responding with faulty information.
However, the application developer and clients can detect (though not attribute)
such faulty behavior.
We assume developers will choose services known for availability.
While we do not discuss it in this paper, Byzantine fault tolerant variations of
private information retrieval~\cite{devet2012optimally,goldberg2007improving}
can be used for better availability guarantees at the cost of higher overhead.
Adversarial clients can also degrade service through denial-of-service attacks.

\paragraph{Intersection attacks}\label{sec:background:intersection}

An ideal private messaging system would make an honest user's actions 
perfectly indistinguishable from all other honest users actions.
Even such a perfect system would still have its limitations;
for example, if only two honest users ever participate, then 
an attacker can infer that those two users are likely to be communicating with each other.

In practice, most private messaging systems do not achieve ideal hiding;
instead they hide a user's actions with a smaller anonymity set,
e.g., the set of all online users, or online writers, or even a smaller subset thereof.
These compromises are made due to practical constraints or the desire for better performance.
For example, if users are allowed to go offline without causing the rest of the system
to stop functioning, then the anonymity set of any active user is necessarily reduced to only
those users active at the same time.
Since these anonymity sets change over time (e.g., based on when users come online),
users may be vulnerable to 
intersection attacks~\cite{danezis2004statistical,kedogan2002limits,mathewson2004practical}
in which the adversary observes a particular user's anonymity sets over time,
and, by intersecting those sets, narrows down the user's overall anonymity set.

\name provably provides access sequence indistinguishability,
but, as with most private messaging systems~\cite{van2015vuvuzela,angel2016unobservable,kwon2016riffle,angel2018,kwon2017atom,tyagi2017stadium,lazar2018karaoke,kwon2019xrd},
for practical reasons allows users to go offline.
Thus, an honest user is hidden amongst all honest online users.
This offers stronger privacy than systems based on k-anonymity.
In such systems, for performance reasons, $k$ is usually set to be much 
smaller than the set of all online users (e.g., $k$ might be the number
of users actively writing in a round).  Furthermore, in each round 
of communication, the chosen set of $k$ users may vary, and so the
adversary has many opportunities to perform intersection attacks on these
small sets.

\subsection{Assumptions}\label{sec:background:assumptions}

To provide access sequence indistinguishability, Talek relies on several assumptions. We assume that server storage capacity is scaled to the number of clients.
We assume that communicating clients already know each others' long-term public keys.
\systemname{} is compatible with bootstrapping keys from other applications~\cite{keybase,DBLP:journals/iacr/DemmlerRRT18} or
identity-based encryption~\cite{boneh2001identity,boldyreva2008identity,lazar2016alpenhorn}.
We assume each server has a public-private key pair, $pk,sk$, generated
using an algorithm $PKGen()$,
and that all server public keys are known to all users.
Establishing such keys is orthogonal to the properties \systemname{} provides.

As with prior work~\cite{angel2016unobservable,angel2018}, \systemname{} supports groups of mutually trusting users
that protect shared secrets.
If this trust is misplaced, the writer's anonymity may be lost,
but PIR still preserves reader anonymity.

 \section{Design Overview} \label{sec:design} \label{sec:overview}

In this section, we give a brief overview of \name's key design decisions.
Later, we present \name's core operations more formally and in greater detail (\S\ref{sec:design:core}),
followed by two optimizations to improve read performance: 
serialized PIR (\S\ref{sec:serialized}) and private notifications (\S\ref{sec:design:notifications}).

\name's core abstraction is a private log,
which enables a single writer to share messages with many readers.
Groups can then be formed with separate logs for every writer,
such that all members subscribe to the logs of all other group members.

In general,
we do not expect to see a single universal \name service;
instead, we expect developers (or federations thereof) will stand up separate \name instances
(e.g., one for Instagram and one for Twitter).
In practice, a single application might even run two parallel instances of \systemname{},
e.g., one for text-based data, and one with higher latency for images.
Running separate instances will allow developers to better
tune \name's parameters to their application.

\subsection{\name's Client Interface}\label{sec:overview:client}

\begin{figure}[tb]
  \begin{center}
    \includegraphics[width=0.30\textwidth]{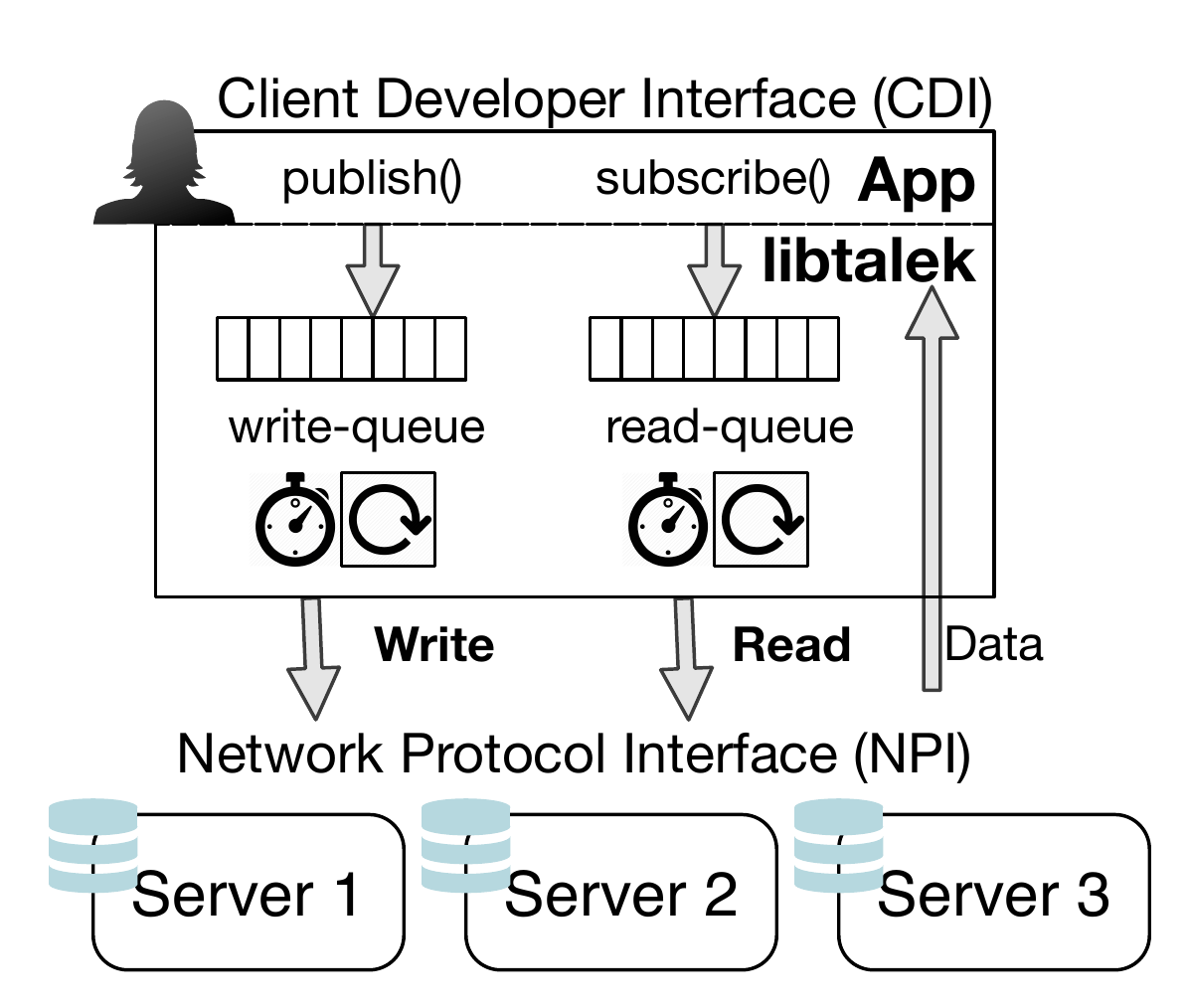}
  \end{center}
  \vspace{-0.8em}
  \caption{\footnotesize \textbf{\name Client Interface.}
    Application calls are translated by the client library into scheduled messages
    with equal-sized parameters and contents that appear random to an adversary.
    Clients behave identically from the perspective of any $l-1$ servers.
  }
  \label{fig:architecture}
  \end{figure}
  
\name achieves our security goal by \emph{requiring all users to behave identically
from the perspective of any colluding set of $l-1$ servers}.
A key step in this direction is decoupling the \emph{real}
rate at which a user generates application read/write requests
from the rate at which network read/write operations are performed.

As Figure~\ref{fig:architecture} illustrates, 
developers link their messaging application to the \systemname{} client library, which
 maintains internal read- and write queues. Whenever a request is issued by the messaging application,
\systemname{} places its payload on the corresponding internal queue.
The \name client library issues equal-sized \texttt{Read} and \texttt{Write} network requests at a rate that is independent of
the user's real request rate, issuing a dummy request (i.e., a $\mathit{FakeWrite}$ or a $\mathit{FakeRead}$) if the respective queue is empty. 
To preserve privacy, a $\mathit{FakeWrite}$ ($\mathit{FakeRead}$) request, including its parameters and payload,
must be indistinguishable from a $\mathit{RealWrite}$ ($\mathit{RealRead}$) request.
\name achieves this by defining a globally-fixed message size, $z$, 
to which messages are split and padded to fit.
Messages are then encrypted with a symmetric authenticated encryption scheme~\cite{bellare1998relations}
to provide confidentiality and authenticity.

\paragraph{Choosing appropriate network rates} In practice, the developer should measure real global usage and sample randomly from this aggregate distribution.
They could also configure a fixed burst of messages every time a user comes online.
Privacy is preserved as long as the distribution of requests is independent of user activity.

\subsection{Oblivious Logging Overview}\label{sec:overview:logging}
Decoupling the real rate at which a user generates read/write requests from the network request rate is insufficient for privacy, as it fails to hide which users read each other's messages.
Hence, \systemname{} uses PIR (\S\ref{sec:background}) to allow users to retrieve records privately. However, PIR alone does not explain how to write messages to the servers or how to let communication partners know which messages to read.

To allow clients to read \emph{and} write messages privately without explicit coordination,
\name issues writes to pseudo-random locations on each server, similar to prior work~\cite{angel2016unobservable,angel2018,van2015vuvuzela,tyagi2017stadium,lazar2018karaoke}.
We call our version of this technique \emph{oblivious logging}.
The sequence of write locations is determined by
applying a pseudorandom function, $PRF$, to a secret log handle shared
between the log's writer and readers.
Readers use PIR to retrieve messages by following the PRF-derived sequence
of locations.

Hence, server simply see a series
of seemingly random writes and PIR-protected reads,
hiding the communication patterns between writers and readers.
Exposure of the log handle (e.g., by an untrustworthy group member)
will expose the writer's access pattern, but not reader consumption patterns,
since the latter are protected information theoretically by PIR's guarantees.

\subsection{\name's Server Design}\label{sec:overview:server}
Servers store a limited set of messages
to allow asynchronous senders and receivers to be decoupled in time.
Since the cost of PIR operations scales linearly with database size,
for good performance we fix the number of messages stored on each server to $n$,
garbage collecting the oldest.
$n$ is directly related to the time-to-live, $TTL$, for messages,
which dictates how tightly synchronized senders and receivers must be.
As the number of clients in the system grows, the system requires larger
values of $n$ to support the same $TTL$.

\begin{figure}
\begin{center}
  \includegraphics[width=0.25\textwidth]{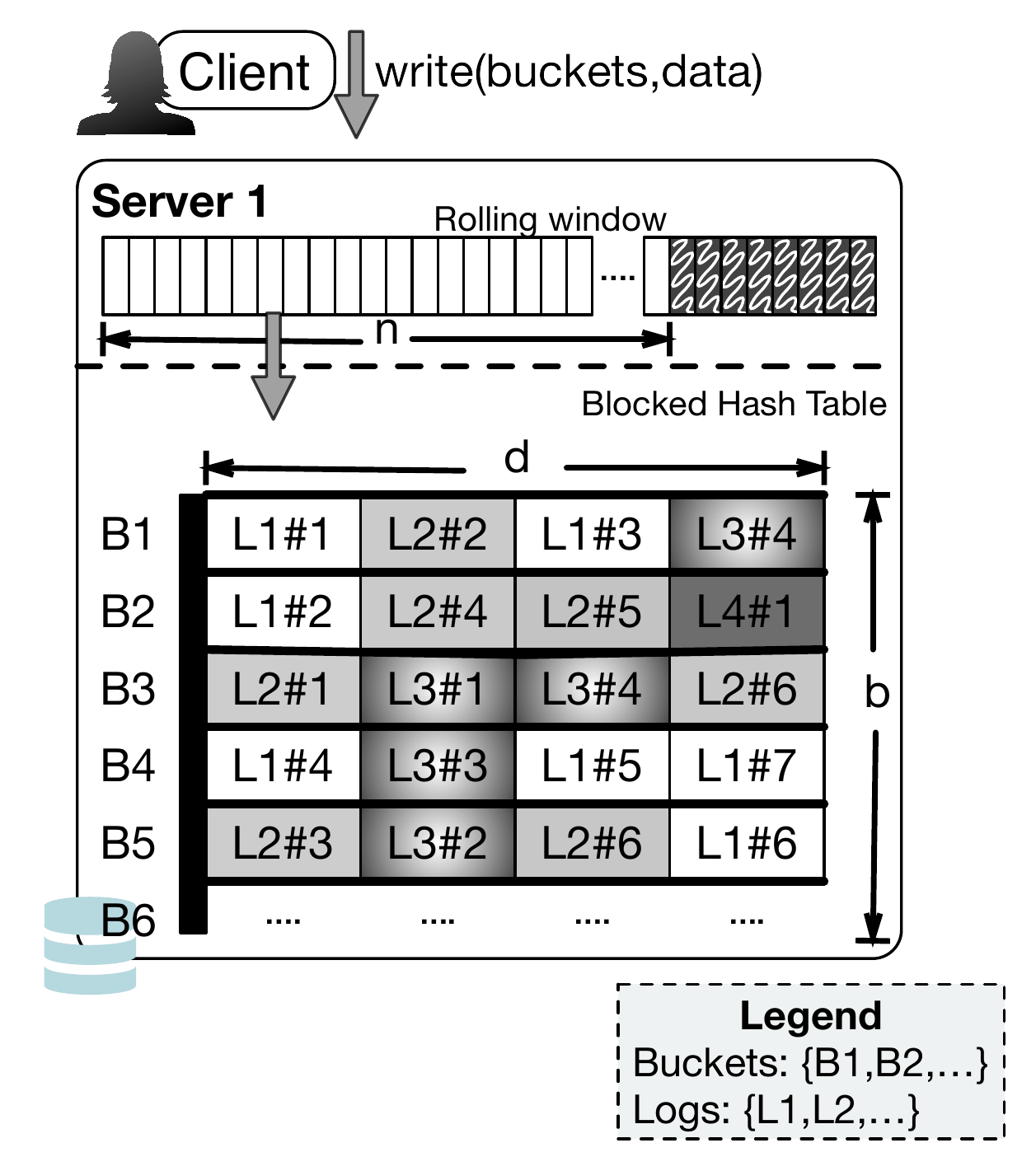}
\end{center}
  \vspace{-1em}
  \caption{\footnotesize
    \textbf{Server Data Structures and Workflow.}
    Garbage collection discards all but the latest $n$ messages.
    Client writes specify two random buckets in which each message can be placed,
    forming a blocked cuckoo hash table.
    Logs are spread over the hash table, and read using PIR\@.
    Messages in the same log are colored with the same shade in the diagram.
  }
\label{fig:obliviouslogging}
\end{figure}

Like prior messaging systems~\cite{angel2018,angel2016unobservable}, 
we face the challenge of finding a way to efficiently pack messages into a
dense data structure that is compatible with PIR. 
One of \name's key innovations is the use of a
blocked cuckoo hash table~\cite{pagh2001cuckoo,dietzfelbinger2007balanced},
characterized by a fixed value of $b$ buckets each containing $d$ messages.
Each client \texttt{Write} request explicitly specifies (to the servers)
two pseudo-randomly chosen buckets in which the message can be inserted.
The servers apply the standard cuckoo-hash insertion algorithm,
which potentially results in cuckoo evictions if both buckets are full, but guarantees that every item
ends up in one of its two hash-selected locations.  
When the client performs a \texttt{Read} request,
the hash table is treated as a PIR database with each hash bucket as an entry.
The client uses PIR to retrieve an entire hash bucket
without revealing to the server which bucket it retrieved.

Blocked cuckoo hashing has a number of desirable properties.
Compared to chained hash tables, all buckets have an equal fixed size,
a necessary requirement for PIR\@.
Further, only a single copy of each message written is stored,
minimizing storage overhead.
To handle collisions, the size of the table does need to be larger than $n$,
but with blocked cuckoo hashing the overhead is generally less than 20\% for
reasonable values of the bucket size $d$.
This also helps minimize the cost of PIR operations, 
which grow with the size of the database.
Finally, there are only two possible locations for each message, and thus a client issues at most two \texttt{Read} requests
to check both buckets where a message could be stored;
this is important as PIR operations are expensive.
If the client finds the message it is looking for in the first bucket,
then it can use its next \texttt{Read} request for another $\mathit{RealRead}$,
rather than querying the second cuckoo hash location.
From the server's perspective,
the client is simply issuing a stream of opaque PIR requests.

 \section{\systemname's Core Operations}\label{sec:design:core}

This section expands on the overview from \S\ref{sec:overview}
to provide details on the \name protocol for creating
single-writer, multi-reader logs.
Log handles (\S\ref{sec:design:logging:handles}) allow readers to find the
latest content from writers without explicit coordination.  
They dictate how messages are written into the servers' cuckoo hash tables (\S\ref{sec:design:cuckoo}),
and how readers retrieve messages using PIR (\S\ref{sec:design:pir}).
Because reads are done with PIR, many readers can poll the same log
repeatedly without revealing information.
Finally, we 
provide some intuition (\S\ref{sec:security:intuition}) for \name's proof of security (Appendix~\ref{sec:security:proof}).

Figure~\ref{fig:state} summarizes the client/server interfaces and state.

\begin{figure}[tb]
\begin{center}
    {\footnotesize
  \begin{tabular}{|l|}
    \hline
    {\bf Client Developer Interface (CDI)} \\
    \texttt{Publish}(log, message)  \\
    \texttt{Subscribe}(log) \\
    \hline
    {\bf Network Protocol Interface (NPI)} \\
    \texttt{Write}(bucket1, bucket2, encryptedMsg, interestVector) \\
    \texttt{Read}(requestVectors[]) $\rightarrow$ encryptedData \\
    \texttt{GetUpdates}() $\rightarrow$ globalInterestVector \\
    \hline
    {\bf Client State} \\
    $\bullet$ $logs$ - List of subscribed logs \\
    $\bullet$ $writeQueue$ - Queue of write operations \\
    $\bullet$ $readQueue$ - Queue of read operations \\
    \hline
    {\bf Server State} \\
    $\bullet$ $log$ - Global log of write operations \\
    $\bullet$ $table$ - Blocked cuckoo hash table \\
    \hline
  \end{tabular}
  }
\end{center}
  \vspace{-0.4em}
\caption{\footnotesize
   \textbf{\systemname{} Interfaces and Client/Server State.}
   Global interest vectors are described in \S\ref{sec:design:notifications}.
}
\label{fig:state}
\end{figure}

\subsection{Notation}\label{sec:notation}

\begin{figure}
\begin{center}
{\footnotesize
\begin{tabular}{cll}
  \multicolumn{3}{c}{\bf Globally Configured} \\
  \toprule
  $l$       & constant          & Number of servers\\
  $n$       & constant          & Number of messages stored on server  \\
  $b$       & constant          & Number of server-side buckets \\
  $d$       & constant          & Depth of a bucket \\
  $z$       & constant          & Size of a single message \\
  $w$       & constant          & Per-user rate of writes \\
  $r$       & constant          & Per-user rate of reads \\
    \multicolumn{3}{c}{\bf Dynamically Measured} \\
  \toprule
  $m$       & variable          & Number of online clients \\
  $TTL$     & $n/(m*w)$         & Lifetime of a message on the server \\
  $load$    & $n/(b*d)$         & Load factor of the server hash table \\
\end{tabular}
}
\end{center}
    \vspace{-0.4em}
\caption{\footnotesize
  \textbf{\name Variables.}
      Globally configured parameters are fixed across all clients and servers
  for an instance of \systemname{}.
  }
\label{fig:parameters}
\vspace{-1.2em}
\end{figure}

For convenience, Figure~\ref{fig:parameters} summarizes constants that parameterize \name's design.

We write $PKEnc_{pk}(text)$ for the encryption of $text$ under $pk$, and $PKDec_{sk}(cipher)$ for the decryption of $cipher$ under $sk$.
Clients also have access to an efficient symmetric encryption scheme.
We write $Enc_k(text)$ for the encryption of $text$ with key $k$, and $Dec_k(cipher)$ for the decryption of $cipher$.
Let $PRF(key, input)$ denote a pseudorandom function family
and $PRNG(seed)$ denote a cryptographically secure pseudorandom number generator.
We use $|$ to denote tagged concatenation.

\subsection{Log Handles}\label{sec:design:logging:handles}

When a user creates a new log, \systemname{} generates a secret \emph{log handle}, $\tau$,
containing a unique ID, $id$, encryption key, $k_{enc}$, and two seeds, $k_{s1}$ and $k_{s2}$.
The log handle is shared between the writer and readers of a log.
All messages in the log are encrypted with $k_{enc}$, e.g., 
$Enc_{k_{enc}}(message)$.
We further assign all messages in a log a sequence number, $seqNo$.
This sequence number is internal to \systemname{}, and does not preclude
application-layer logic to track and order messages.
The two seed values are used in conjunction with a
pseudorandom function family, $PRF(seed, seqNo) \in \{0 \ldots (b-1)\}$,
to produce two unique and deterministic sequences of bucket locations for writes.
Similar to frequency hopping~\cite{ephremides1987design,dixon2008phalanx}
and constructions in prior work~\cite{angel2016unobservable,angel2018,van2015vuvuzela,tyagi2017stadium,lazar2018karaoke},
log handles allow writers and readers to agree on a pseudorandom sequence of buckets
without online coordination. 
As discussed in \S\ref{sec:problem},
additional properties, like forward secrecy,
can be layered atop this basic construction.

\subsection{Writing into a Cuckoo Hash Table}\label{sec:design:cuckoo}

A key part of \name's design is that server-side state is arranged in a
blocked cuckoo hash table~\cite{pagh2001cuckoo,dietzfelbinger2007balanced},
where each server's storage is organized into $b$ buckets,
each bucket storing $d$ messages, each of size $z$.
The value of $b$ is chosen in conjunction with the PRF family,
such that the output of a PRF is uniformly distributed across the buckets.
The servers' message capacity, $n$, is chosen as a fraction of the capacity of the cuckoo table,
$b \cdot d$. This fraction is set to ensure with high probability
that a message will fit with minimal rearranging of the cuckoo table~\cite{dietzfelbinger2007balanced};
hence insertion costs $O(1)$.

Because cuckoo hashing is history-dependent and PIR requires
consistent replicas across all servers, 
we must ensure that all writes are inserted in the same order.
We address this via serialized PIR (\S\ref{sec:serialized}).
The algorithm for inserting into a blocked cuckoo hash table
is also randomized, so the servers agree on a random seed, $s_{cuckoo}$,
which they use to ensure they all make the same (pseudo-)random choices
(if one or more servers make different choices, it affects liveness, but not privacy).

When the \name client library dequeues a message, $M$, with sequence number $seqNo$
from its internal write queue, it runs the following $\mathit{RealWrite}(\tau, seqNo, M)$ algorithm: 
  \begin{align*}
    \beta_1 &= PRF(\tau .k_{s1}, seqNo) \in [0..b-1]\\
    \beta_2 &= PRF(\tau .k_{s2}, seqNo) \in [0..b-1]\\
    data    &= Enc_{\tau .k_{enc}}(seqNo | M)
  \end{align*}
and sends the \texttt{Write} request, $\beta_1|\beta_2|data$, to the servers.  
If the write queue is empty, it instead 
runs $\mathit{FakeWrite()}$; i.e., it  
chooses $\beta_1$ and $\beta_2$ at random in $[0..b-1]$
and computes 
  \begin{align*}
    data    &= Enc_{\tau .k_{enc}}(dummySeqNo | 0^z)
  \end{align*}
i.e., the encryption of a reserved sequence number and $z$ zeroes.  

Upon receiving the \texttt{Write} request, 
each server follows the standard cuckoo insertion algorithm:
\squishenum{}
  \item Delete the $n$-th oldest element.  \item Insert $\beta_1|\beta_2|data$ into either bucket $\beta_1$ or bucket $\beta_2$
    if there is spare capacity in either bucket.
  \item If both buckets are full, choose $\beta_e \in \{\beta_1, \beta_2\}$,
    using randomness derived from $s_{cuckoo}$.
    Let $\delta_e=\beta_1|\beta_2|data$.
  \item Repeat the following until all values are inserted
    \squishenum{}
      \item Try to insert $\delta_e$ in $\beta_e$ if the bucket has space.
      \item If not, randomly evict an entry in $\beta_e$ and insert $\delta_e$ there.
      \item Set $\delta_e$ to be the evicted value, and set $\beta_e$ to its alternate bucket location.
    \squishenumend{}
\squishenumend{}

\subsection{Reading via PIR}\label{sec:design:pir}

When the \name client library dequeues a log read at sequence number $seqNo$
from its internal read queue, it generates a \texttt{Read} request using the following
$\mathit{RealRead}(\tau, seqNo)$ algorithm.
The library performs a PIR read (following the protocol from \S\ref{sec:background})
for bucket $PRF(\tau .k_{s1}, seqNo)$ (i.e., the first possible cuckoo location).
The library attempts to decrypt every message in the bucket by computing $Dec_{\tau .k_{enc}}(data)$.
If decryption succeeds and matches $seqNo$, $M$ is returned to the application.
If not, the next time a \texttt{Read} request should be issued (as determined by the randomized
reading schedule), the library performs a PIR read for the second cuckoo location,
i.e., $PRF(\tau .k_{s2}, seqNo)$.

If the internal read queue is empty,
the library runs $\mathit{FakeRead()}$, which issues a PIR read for an arbitrary bucket.

\subsection{Control Logs}
\label{sec:control}
To facilitate control messages between users, we 
establish a \emph{control log} between every pair of users who want to communicate.
We expect the log handle for the control log to be generated and exchanged out of band
when users exchange and verify public keys.

When a user wants to give log access to one of their friends,
they send the new handle via the control log shared with the intended subscriber.
Once all subscribers have access to the log handle, 
the log writer can write messages once to be read by many subscribers,
instead of sending individual messages through each pairwise control log.
Revocation occurs by creating a new handle and sharing it with the non-revoked 
subscribers.

Control logs are also used by \systemname{} to coordinate between users.
If a user has been offline for an extended period, she can ask the writer for the most recent
sequence number of a particular log. Similarly, users can send retransmission requests for missed messages.
By definition, these requests leak information to the sender,
but the sender and receiver are assumed to trust each other.
A client also uses a control log to periodically send heartbeat messages to itself.
If these messages are lost, it serves as a hint to the client of a denial-of-service attack.

\subsection{Security Analysis}
\label{sec:security:intuition}
We formally prove the security of oblivious logging in \S\ref{sec:security:proof}.
Informally, we assume that the writer trusts the consumers of its log
and prove security by reduction to cryptographic assumptions.
For \texttt{Read}, we rely on the security properties offered by PIR\@.
PIR queries that correspond to legitimate requests are indistinguishable from a PIR query for a random item~\cite{chor1998private}.
For \texttt{Write}, we rely on the security properties of a PRF and our encryption algorithm.
We use a symmetric authenticated encryption algorithm for message payloads.
For any \texttt{Write}, the bucket locations are either generated by a PRF using the
log handle's seed values, $(k_{s1},k_{s2})$, or chosen at random.
An attacker who can distinguish between these two cases immediately breaks the PRF's security.

A malicious client has the ability to impact service availability by deviating
from the protocol and sending all its writes to a single bucket.
This attack (and other forms of denial-of-service)
is limited by servers enforcing the fixed \texttt{Write} rate $w$ per client,
the number of Sybil clients the attacker can obtain, and the size of the database, $n$.
The attack is further mitigated by
the self-balancing nature of cuckoo tables, where legitimate
messages can be evicted to their alternate locations, and 
the fact that the malicious client will not know which buckets to target to
disrupt a specific communication.

 \section{Serialized PIR}\label{sec:serialized}

Similar to Riffle~\cite{kwon2016riffle}, 
to reduce the network costs of \texttt{Read} requests for the client, 
we introduce a \emph{serialized} version of IT-PIR,
which offloads work from clients to the servers.
Unlike Riffle's design,
ours avoids the need for the servers to keep per-client state.

At a high level, 
we choose an arbitrary server to be the \emph{leader}, $\mathcal{S}_0$, with the
rest of the servers forming the \emph{follower} set, $[\mathcal{S}_1, \ldots, \mathcal{S}_{l-1}]$.
All \texttt{Read} and \texttt{Write} requests are directed to the leader.
The leader adds a global sequence number to each request 
(to ensure operations are handled consistently across the servers)
and forwards it to the followers.
Each \texttt{Read} request sent to the leader contains a PIR query 
for each server, encrypted under its respective public key.
The followers each send their response back to the leader (rather than the client).
The response is carefully masked 
to preserve the confidentiality of each server's results
while allowing the leader to combine them on behalf of the client.
Overall, the client's outbound request bandwidth is the same as before,
but the inbound bandwidth is only the size of a single response,
rather than $l$ responses.

To issue a serialized PIR request
the client, $\mathcal{C}$ proceeds as follows.
\squishenum{}
  \item $\mathcal{C}$ generates $b$-bit PIR requests
    for each server, $\{q_0, \ldots, q_{l-1}\}$
    ($b$ is the number of server hash buckets).
  \item $\mathcal{C}$ generates a high-entropy random seed for each server, \\
    $\{p_0, \ldots, p_{l-1}\}$
  \item $\mathcal{C}$ encrypts each server's parameters with its respective public key and generates
    a \texttt{Read} request,  \\
    $PKEnc_{pk_0}(q_0|p_0), \ldots, PKEnc_{pk_{l-1}}(q_{l-1}|p_{l-1})$
  \item $\mathcal{C}$ sends this request to the leader, $\mathcal{S}_0$, who forwards it to
    the followers along with a global sequence number.
  \item In parallel, each server, $\mathcal{S}_i$, decrypts its respective PIR request vector, $q_i$
    and computes its response, $R_i$.
  \item Each server, $\mathcal{S}_i$, also computes a random one-time mask, $P_i=PRNG(p_i)$, from the seed parameter.
    This mask should have the same size as $R_i$.
  \item Each server, $\mathcal{S}_i$, responds to $\mathcal{S}_0$ with $R_i \bigoplus P_i$.
  \item $\mathcal{S}_0$ combines the server responses and responds to $\mathcal{C}$ with
    $R_0 \bigoplus P_0 \bigoplus \ldots R_{l-1} \bigoplus P_{l-1}$
  \item $\mathcal{C}$ restores the bucket of interest by XOR'ing this response
    with each server's mask,
    $P_0 \bigoplus \ldots \bigoplus P_{l-1}$
\squishenumend{}

\vspace{0.04in}
\noindent {\bf Security:}
With respect to privacy, this serialized variant of PIR is provably equivalent
to the traditional PIR scheme described in \S\ref{sec:background:pir}.
The proof is straightforward and sketched in \S\ref{sec:security:serializedpirproof}.
Hence, even if the leader is malicious,
it can only undermine liveness, not privacy.

\vspace{0.04in}
\noindent {\bf Correctness and Liveness:}
The leader is only responsible for assigning a global sequence number,
which does not affect security or correctness.
If the leader misrepresents the global sequence number of a message,
it could cause those replicas to become inconsistent.
Because any follower could also deny service by failing to respond or deviating from the protocol,
the leader is in no more privileged a position
to affect correctness or liveness of the system than any other server in the system.
Furthermore, serialized PIR is compatible with the Byzantine-fault-tolerant varieties of PIR, which can be used to improve liveness guarantees (\S\ref{sec:background:availability}).

 \section{Private Notifications}
\label{sec:design:notifications}

Regularly polling for new messages presents two problems.
First, because every user polls at the same rate, message latency increases as a user subscribes to more logs.
Second, it is hard to know which log to poll at any
given time.

Inspired by previous uses of Bloom filters for private membership queries~\cite{erlingsson2014rappor,chang2005privacy},
we introduce a private notification system
that allows users to determine when new messages have been
published to a log without revealing the list of logs to which they subscribe.
By detaching reads from notifications, clients can prioritize reads
and reduce how often they read buckets for logs without new messages.

The private notifications system works as follows:
\begin{enumerate}
\item With every \texttt{Write} request, clients send an \emph{interest vector}, privately encoding the log handle and message sequence number of this request.
\item Servers maintain a \emph{global interest vector} which encodes the set of (log handle, message sequence number)-pairs for all messages on the server.
\item Clients periodically query the lead server for the global interest vector.
\end{enumerate}

\paragraph{Computing an Interest Vector}
During a $\mathit{RealWrite}(\tau , seqNo, M)$, the client creates an interest vector by 
inserting into an empty Bloom filter that uses \emph{cryptographic} hash functions:
\begin{align*}
    intVec \leftarrow BloomFilter()\\
    intVec.insert(\tau .id|seqNo)
  \end{align*}
For a $\mathit{FakeWrite}$ request, we insert a random value instead of $\tau .id|seqNo$.

\paragraph{Maintaing a Global Interest Vector}
The global interest vector is the union
of the interest vectors written to the server.
Servers periodically sign their global interest vector
and exchange signatures in a multi-signature scheme~\cite{boneh2003aggregate},
which allows the lead server to aggregate all signatures into a single compact one.
Upon a $\mathit{GetUpdates()}$ request from the client, the leader sends that signature, along with the global interest vector, 
to the client for verification.

In practice, \name employs compressed Bloom filters~\cite{compressedbloom},
and both clients and servers maintain a window of Bloom filter deltas~\cite{compressedbloom},
with each delta summarizing changes since the previous delta.  
By discarding old deltas as new ones arrive, 
we avoid saturating the global interest vector.
Altogether, compressed Bloom filters with delta compression result in updates of less than 10k bytes for our experiments with 1M messages stored on the server. We set clients to fetch updates every
20 reads to further amortize this cost.

\paragraph{Security Analysis}
The security of private notifications relies on the cryptographic
hash functions (modeled as a random oracle) used in the Bloom filter.
As long as we use a log ID with sufficient entropy,
each interest vector provides a negligible advantage in the indistinguishability security game.
We give a formal proof in \S\ref{sec:security:proof} (Game 4).

Private notifications are only used to prioritize reads on the internal request queue.
As such, it has no impact on \name's security goals,
since it has no visible effect on the network protocol interface.
It simply reorders the internal schedule of private requests.

 \section{Implementation}
\label{sec:implementation}
To evaluate \systemname{}'s practicality,
we have implemented a prototype in approximately 6,200 lines of code; the source code is available on GitHub.We have implemented two versions. The first, written in Go, runs entirely on the CPU\@.
The second offloads PIR operations to the GPU using a kernel written in C on OpenCL\@,
sharing memory between the CPU implementation and the GPU\@.
The prototype uses SipHash~\cite{aumasson2012siphash} as the PRF,
and NaCl's box API~\cite{bernstein2012security} for public and symmetric
authenticated encryption.  NaCl's API
relies on a combination of Curve25519, Salsa20, and Poly1305.

 \section{Evaluation}
\label{sec:evaluation}
\FloatBarrier{}

Our evaluation addresses the following questions:
\squishenum{}
  \item \textbf{\ref{sec:eval:costperop}} What is the cost of operations for clients and servers?
  \item \textbf{\ref{sec:eval:cover}} What is the cost of cover traffic?
  \item \textbf{\ref{sec:eval:throughput}} How does system performance scale with more users?
    \item \textbf{\ref{sec:eval:comparison}} How does \systemname{} compare with previous work?
    \squishenumend{}

\subsection{Setup}
All experiments are conducted on Amazon EC2 P2 instances. These virtual machines are allocated 4 cores on an Intel Xeon E5--2686v4 processor
and 61 GB of RAM\@.
They also include an NVIDIA K80 GPU with 2496 cores and 12GB of memory.
We use 3 servers;
one is chosen as the leader and the others are followers.
We allocated two VMs to run user clients.
Each user client issues periodic \texttt{Read} and \texttt{Write}
requests to the server.

While our experiments are run in a single data center,
we expect the performance to be similar for a more
realistic cross-data center setting.
This would incur higher network latency, both to reach the leader
and in communicating between servers.
However these latencies will not impact the results here which
focus on the main bottleneck:
the server-side computational cost of \systemname.

To evaluate a realistic workload,
we used the Ubuntu IRC logs from 2016~\cite{ubuntu-irc},
consisting of 1554790 messages over 32,834 unique usernames.
We generate a unique log in \systemname{} for each writer in an IRC channel.
We varied the number of users, $m \in (0, 32K]$,
and the number of messages in the database, 
$n \in \{10\text{K}, 32\text{K}, 100\text{K}, 500\text{K}, 1\text{M}\}$.
We fix the message size to $z=1KB$.
In order to understand the trade-off between
bandwidth and end-to-end message latency,
we vary the global client read and write rates.

\subsection{Cost of Operations}
\label{sec:eval:costperop}

\begin{figure}[tb]
\begin{center}
{\footnotesize
\begin{tabular}{l r r r}
   & \multicolumn{3}{c}{\bf Messages on Server ($n$)} \\
   & \multicolumn{1}{c}{10K}      & \multicolumn{1}{c}{100K}  & \multicolumn{1}{c}{1M} \\
  \toprule
  {\bf Client CPU costs ($\mu$s)}     & & & \\
  Generate new log handle         & 7753  & 7753  & 7753  \\
        Write                           & 67    & 67    & 67 \\
      Issue PIR query                 & 65    & 574     & 6888  \\
      Process PIR response            & 146  & 146  & 146  \\
      
  \toprule
  {\bf Server CPU costs (ms)}     & & & \\
        PIR Read: CPU                   & 1.34        & 11.10        & 88.10  \\
            PIR Read: GPU                   & 0.07       & 0.54        & 4.36  \\
            Write                           & 0.02  & 0.02  & 0.02  \\        
  \toprule
  {\bf Server storage costs (MB)} & & & \\
    1 KB messages                   & 24        & 241         & 2410  \\

  \toprule
  {\bf Network costs (KB)}             & & & \\
    GetUpdates                      & 0.21        & 1.40        & 14.00  \\
      Read request                    & 0.96        & 9.39        & 93.72  \\
    Read response                   & 4.16  & 4.16  & 4.16  \\
          Write request                   & 1.08  & 1.08 & 1.08 \\
    
  \bottomrule
\end{tabular}
}
\end{center}
  \vspace{-0.4em}
  \caption{\footnotesize
    \textbf{Cost of Individual \name Operations.} 
    We vary the number, $n$, of 1KB messages stored on the server.
          }
\label{fig:eval:costperop}
\end{figure}

To understand \systemname's costs, we benchmark different components of
the system. Each value is the average of 200 runs.
We vary the number of messages on the server, $n \in \{10\text{K}, 100\text{K}, 1\text{M}\}$.
We fix the bucket depth in the blocked cuckoo table to 4,
such that clients retrieve 4 messages at a time.
This depth allows the cuckoo table to support a load factor of 95\%.
The number of buckets is chosen to hold $n$ messages at
the maximum load factor for the table.
Figure~\ref{fig:eval:costperop} highlights the results.

In general, client costs are low due to IT-PIR\@.
Each \texttt{Write} encrypts the message and uses a PRF to determine the bucket location.
The cost of generating a PIR query for \texttt{Read} increases with the database size.
Larger values of $n$ translate to more buckets and larger PIR request vectors.

For the server, we implement two versions of IT-PIR\@.
Our CPU implementation streams the database through the local CPU cache while accumulating responses
to service each batch of queries. 
Its performance is limited by memory-bus throughput.
The GPU implementation accelerates performance by 1--2 orders of magnitude
by taking advantage of the inherent parallelism of PIR operations across many GPU cores
and the optimized on-device memory hierarchy.
Unless otherwise stated, henceforward, results are for our GPU implementation.
Batch coding or preprocessing~\cite{lueks2015sublinear} would further improve
this throughput bottleneck, by an estimated 3$\times$,
at the cost of higher latency for writes to become visible to subsequent reads.
Writes incur negligible cost compared to the cost of reads.

Storage costs scale as expected, 
since our current implementation stores all messages twice.
Writes are applied to the working copy stored in DRAM\@.
Periodically, a snapshot of this state is copied into the GPU\@.
Read requests are batched and forwarded to the GPU. 
The leader is free to reorder reads without violating serializability.

Network costs between client and server are minimal.
Clients must submit a read request containing a $b$-bit vector for each server.
The size of \texttt{Read} responses and \texttt{Write} requests are
within a small factor of the message size.
The global interest vector returned from \texttt{GetUpdates} 
grows linearly with $n$ in order to preserve a fixed false positive rate of 0.02
and a $TTL$ of 100 write intervals.
This cost is independent of message size.
Updates trade off bandwidth with false positive rate.
The network costs per operation are identical between servers,
as both \texttt{Read} and \texttt{Write} operations simply relay
from the leader to followers.

\subsection{Cost of Cover Traffic}
\label{sec:eval:cover}
\begin{figure}[tb]
  \begin{center}
    \includegraphics[width=.38\textwidth]{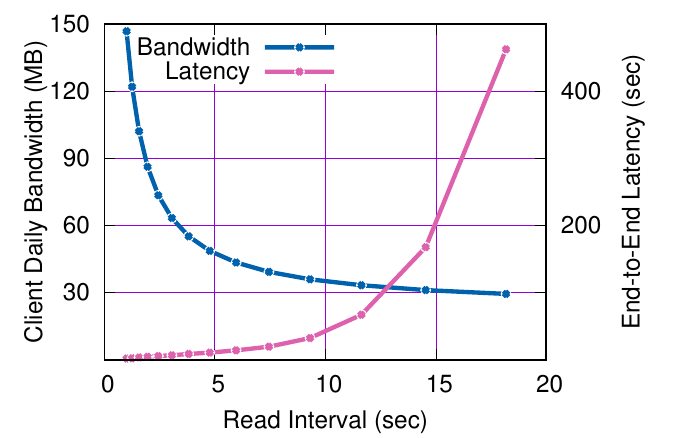}
  \end{center}
  \vspace{-0.4em}
  \caption{\footnotesize
  \textbf{Impact of Read Interval.}
    Average daily bandwidth per client and end-to-end message latency as a function
    of read interval when replaying Ubuntu IRC logs from 2016.
    When clients are configured to send messages once per second,
    each client sends/receives $\sim148MB$ per day to achieve
    end-to-end message latencies $<2$ seconds.
    As we increase the interval of the read schedule, 
    more read requests represent real work, at the expense of
    end-to-end latency.
  }
\label{fig:eval:bandwidthlatency}
\end{figure}

Because each \systemname{} client must generate network traffic on a regular
schedule that is independent from the user's real usage,
developers must choose a global schedule that is appropriate for their application.
Naturally, there is a trade-off between efficiency and message latency.
To quantify this tradeoff, we ran an experiment replaying
the Ubuntu IRC logs into \systemname{}.
We model clients as mobile devices that only send requests when the device is online.
Studies have shown that users are engaged with their mobile devices only $8.6\%$ of the time~\cite{chen2015smartphone}.
We configured servers to store 524,000 messages,
such that servers can keep up with read requests if each client reads one message per second.
We then configure client write schedules to maintain a message TTL of one day on the servers.

Figure~\ref{fig:eval:bandwidthlatency} shows the effect of increasing client
read intervals.
When clients are configured to read every second,
average end-to-end message latency is 1.71 seconds.
The latency exceeds the read interval, 
since many clients subscribe to multiple logs,
so a burst of writes from different writers
may be bottlenecked by the read interval.
This latency comes at the cost of every client using 148MB
of bandwidth per day.
For comparison, the installer for Adobe Reader is $\sim$200 MB, and the top ten most downloaded apps in 2018 on the Google Play store~\cite{appsizes} range from 13-66 MB.
If daily bandwidth usage is of concern, we can drastically reduce it by increasing the read interval by a few seconds,
while keeping latencies below 10 seconds.
At higher read intervals, we reach optimal network usage,
where most read requests corresponds to real work.
However, as read requests are increasingly rate-limited,
message latencies grow quickly.
In practice, application developers will need to choose read intervals
based on acceptable usability and network costs.
This design decision will depend on application workloads.

In future sections, we fix the read interval to 5 seconds,
balancing network and latency considerations.
At this read interval, end-to-end message latencies are 10.7 seconds,
about half of read requests are $\mathit{FakeRead}$ requests, and
each client will send about 49.4MB per day over the network.
For comparison, Snapchat, the \#2 top downloaded mobile app for both iOS and Android,
consumes 14-86MB per day per client~\cite{snapchat-data}.
Assuming the system supports 32,000 active users, servers
will need a network connection of at least 18.3\,MB/s.

\subsection{Throughput}
\label{sec:eval:throughput}
\begin{figure}[tb]
  \begin{center}
    \includegraphics[width=.34\textwidth]{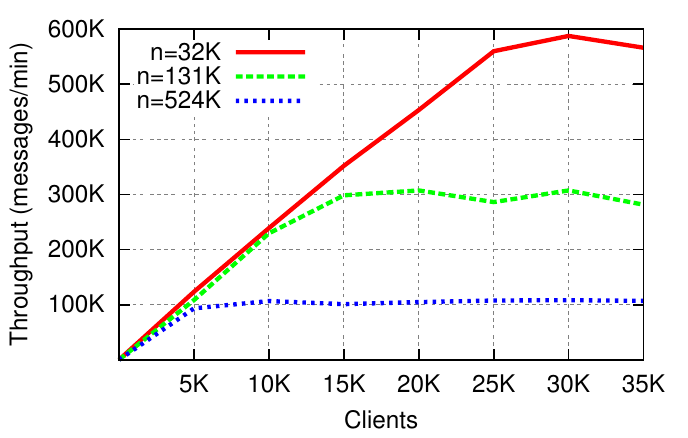}
  \end{center}
  \vspace{-0.9em}
  \caption{\footnotesize
  \textbf{\name's Throughput.}
    Throughput of the system when varying the number of real clients.
    Each client independently issues read and write requests every 5 seconds.
    Each line represents a different value for $n$, the number of messages
    on the server.
    Larger values of $n$ require scanning a larger table,
    resulting in lower throughput.
  }
\label{fig:eval:throughput}
\end{figure}

To understand \systemname's peak performance, we experimented
with a simulated messaging workload. Each client sends a message every five seconds,
and receives a message every five seconds.
For each data point, we spawn a number of clients and measure the leader's
response rate over 5 minutes, giving the system enough time to reach steady-state performance.
Writing in \systemname is cheap, so we limit our workload such that each written message
must be read before being garbage collected.
If writes were not throttled, servers could easily accommodate higher write throughput,
while reads are bottlenecked by PIR computation.

Figure~\ref{fig:eval:throughput} shows the results for three values of 
$n \in \{32\text{K}, 131\text{K}, 524\text{K}\}$,
the number of messages stored on the server.
For small numbers of clients, the server achieves linear growth in throughput,
demonstrating that the PIR operations are keeping up with read requests.
The throughput is bottlenecked by the GPU's PIR process.
Smaller values of $n$ correspond to a smaller cuckoo table,
resulting in cheaper PIR operations and higher throughput.
We only evaluate the system with numbers of clients, $m$, such that $m<n$,
corresponding to a message lifetime of at least one round of reads.

To further improve server performance,
each \name ``server'' could internally consist of multiple machines, 
each handling a portion of the PIR database.
We defer this optimization to future work.

\subsection{Comparison with Prior Work}\label{sec:eval:comparison}

\name explores a new design point in the space of private messaging.
To understand the performance implications,
we compare \name's performance to representatives of other interesting 
design points using their publicly available implementations.
Pung~\cite{angel2016unobservable} and its refinement~\cite{angel2018},
which we dub Pung++, share a similar security goal with \name,
but adopt a stronger threat model incompatible with IT-PIR.
Riposte~\cite{corrigan2015riposte} shares \name's threat model and employs ``reverse'' IT-PIR,
but offers anonymous broadcast (i.e., senders are anonymous but no private read operation is supported)
and targets a weaker privacy goal (k-anonymity).
Vuvuzela~\cite{van2015vuvuzela} shares \name's threat model
but targets another (weaker) privacy goal (differential privacy).
We describe additional protocol differences in \S\ref{sec:related}.

In Figure~\ref{fig:eval:comparecomplexity}, we summarize the asymptotic 
complexity of each system.
These asymptotics, however, hide vastly different constants. 

Hence, Figure~\ref{fig:eval:comparison} shows the concrete single-threaded throughput of each system.
For each system, we sanity checked that our results were consistent with those
reported by the authors.
For Pung, we use a single server.
For \systemname{}, Vuvuzela, and Riposte, systems that rely on an anytrust threat model,
we use a common security parameter of 3 servers. In these systems, larger numbers of servers
only improves security, not performance.
Write performance is similar between \systemname{}, Pung++, and Vuvuzela,
which all dominate Riposte due to its $O(\sqrt{n})$ cost. 
Read performance, however, demonstrates further tradeoffs.
Pung++'s throughput lags \name's CPU implementation by up to two orders of
magnitude, which in turn lags Vuvuzela by 1-3 orders of magnitude.
The Riposte implementation does not include an implementation for reads due to
its focus on broadcast applications.
Pung++ reads are more expensive predominantly due to its use of homomorphic
encryption for C-PIR (to support a stronger threat model) 
and differences in its read protocol. 
Vuvuzela's weaker security goal allows it to scale better than \name.
\name's GPU implementation improves read performance by $\sim 20\times$.

\begin{figure}
\begin{center}
{\footnotesize
\newcommand*\xmark{\ding{55}}
\setlength{\tabcolsep}{3pt}
\begin{tabular}{l|l cccc}
  &              & \systemname{} & Pung++ & Riposte & Vuvuzela \\
    \hline
  {\bf Client CPU}&  Read            & $O(ln)$    & $O(\sqrt{n})$         & \xmark{}              & $O(l)$ \\
   &  Write           & $O(1)$    & $O(1)$              & $O(\sqrt{n})$         & $O(l)$ \\

    \hline
  {\bf Total Server CPU}  &  Read            & $O(ln)$   & $O(n)$      & \xmark{}              & $O(\frac{l^2+ln}{n^2})$ \\
   &  Write           & $O(l)$    & $O(1)$         & $O(ln)$         & $O(\frac{l^2+ln}{n^2})$ \\
  
    \hline
  \multicolumn{2}{l}{\bf Total Server Storage}  & $O(ln)$ & $O(n)$           & $O(ln)$     & $O(l+n)$ \\
  
    \hline
  {\bf Network}  &  Read request    & $O(ln)$   & $O(\sqrt{n})$      & \xmark{}              & $O(1)$ \\
  &  Read response   & $O(d)$    & $O(1)$             & \xmark{}              & $O(1)$ \\
  &  Write           & $O(1)$    & $O(1)$             & $O(l\,\sqrt{n})$      & $O(1)$ \\
  
  \hline
\end{tabular}
}
\end{center}
\vspace{-0.4em}
\caption{\footnotesize
  \textbf{Asymptotic Comparison.} 
  We compare costs between related
work that uses an anytrust or stronger threat model. Parameters are the number of servers, $l$,
the number $n$ of client messages in the system and, for Talek, the number of messages in a bucket $d$.
  Riposte does not specify a read mechanism. 
  Client CPU is for one read/write request. Total server CPU is the total
  cost of $l$ servers (or for Pung++, a single server) for one read or write request. 
  Network costs are for read/write requests between the client and the server(s). Pung++'s costs do not include constructing and sending an oracle that maps message indices to locations within a bucket (\S\ref{sec:related}), while Talek's costs do not include  private notifications, as both are orthogonal to the main design.
}
\label{fig:eval:comparecomplexity}
\end{figure}

 \section{Related Work}
\label{sec:related}

\begin{figure}[tb]
  \begin{center}
    \includegraphics[width=.40\textwidth]{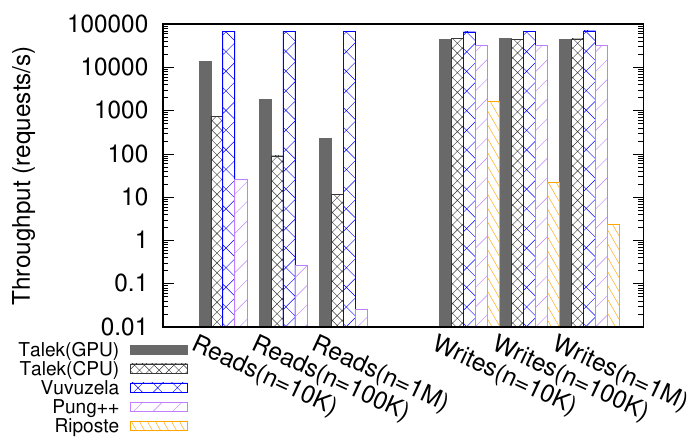}
  \end{center}
  \caption{\footnotesize \textbf{Performance Comparison}
    of \texttt{Read} and \texttt{Write} handlers of various systems.
    Because Riposte is a broadcast protocol, it does not include a read operation.
    Note the log scale on the y-axis.
                  }
\label{fig:eval:comparison}
\end{figure}

\begin{figure}[tb]
\begin{center}
{\footnotesize
\setlength{\tabcolsep}{1.1pt}
\begin{tabular}{l c c c l}
  {\bf System}  & {\bf Security}  & {\bf Threat}  & {\bf Technique}  & {\bf Application} \\
  {\bf }        & {\bf Goal}      & {\bf Model}   & {\bf }            & {\bf } \\
    \toprule
  \systemname{}                           & indisting.  & $\geq$ 1 & IT-PIR & group msg. \\
  Pynchon~\cite{sassaman2005pynchon}      & k-anon.     & $\geq$ 1& mixnet/IT-PIR & email \\
    Riffle~\cite{kwon2016riffle}            & k-anon.     & $\geq$ 1 & mixnet/IT-PIR & file-sharing\\
  Riposte~\cite{corrigan2015riposte}      & k-anon.     & $\geq$ 1 & IT-PIR & broadcast \\
  Dissent~\cite{corrigan2010dissent}      & k-anon.     & $\geq$ 1 & DC-net & broadcast \\
    Atom~\cite{kwon2017atom}                & k-anon.     & $\geq$ $f$ & mixnet & broadcast \\
  Vuvuzela~\cite{van2015vuvuzela}         & diff. privacy & $\geq$ 1 & mixnet & 1--1 msg. \\
  Stadium~\cite{tyagi2017stadium}         & diff. privacy & $\geq$ $f$ & mixnet & 1--1 msg. \\
  Karaoke~\cite{lazar2018karaoke}         & diff. privacy & $\geq$ $f$ & mixnet & 1--1 msg. \\
    Pung~\cite{angel2016unobservable,angel2018}       & indisting.  & 0 & C-PIR & group msg. \\
  ORAM~\cite{stefanov2013multi,stefanov2013oblivistore,stefanov2013path} & indisting. & 0 & ORAM & storage \\
  DP5~\cite{borisov2015dp5}               & indisting.  & $\geq$ 1 & IT-PIR & chat presence \\
  Popcorn~\cite{gupta2016scalable}        & indisting.  & $\geq$ 1 & C-PIR/IT-PIR & video stream \\
  XRD~\cite{kwon2019xrd} 				  & indisting.	& $\geq$ $f$ & mixnet & 1--1 msg. \\
  \bottomrule
\end{tabular}
}
\end{center}
\caption{\footnotesize
  \textbf{Comparison of Privacy Systems.}
  Indistinguishability-based security goals offer the strongest level of privacy.
      Systems based on k-anonymity and differential privacy leak information over time~\cite{danezis2004statistical,kedogan2002limits,mathewson2004practical}.
          The threat model column denotes the number of servers that must be honest for
  security properties to hold; systems marked with $f$ require a fraction (e.g., 20-80\%) of servers to be honest.
  }
\label{fig:related}
\end{figure}

We provide an overview of related systems that hide communication patterns.
Unger et al.\ provide a more detailed survey~\cite{unger2015sok}.
For convenience, Figure~\ref{fig:related} summarizes the discussion.

\paragraph{PIR-based systems}
Theoretical work proves that the collective work required by the servers to answer a client's request scales at least linearly with the size of the database~\cite{beimel2000reducing}. While it is possible to circumvent this lower bound using database preprocessing~\cite{beimel2000reducing} or an offline/online model~\cite{gibbs2019pir}, we cannot take advantage of such approaches since our database changes constantly.

On the systems side, some work in this space leverages assumptions about specific application
workloads to make PIR practical in a particular setting.
For example,
DP5~\cite{borisov2015dp5} is a private chat presence system.
The security of the protocol depends on chat presence workloads
and does not generalize to group messaging.
Similarly, Popcorn~\cite{gupta2016scalable} uses both C-PIR and IT-PIR
to construct a private read-only video streaming system over a static video database.

Some systems support general-purpose anonymous writing,
but do not support private reading. Ostrovsky and Shoup propose 
private information storage (PIS)~\cite{Ostrovsky:1997:PIS:258533.258606},
which allows a client to write to a row in a database of $n$ rows 
without revealing which row was updated.
Writing via PIS is expensive, incurring poly-logarithmic communication,
compared with $O(1)$ for \name.
Riposte~\cite{corrigan2015riposte} expands on this work to support
a scalable broadcast system.
Riposte and \systemname{} share a similar anytrust threat model,
but Riposte has a weaker security goal based on k-anonymity within a round of
communication. Riposte does not promise privacy over multiple rounds of communication,
and writes require $O(\sqrt{n})$ messages. 
Other systems, like \name, aim for general-purpose private messaging,
and hence support a wider range of applications.
For instance, Pynchon Gate~\cite{sassaman2005pynchon} is system where
emails are sent to servers via mixnets.
Emails are dumped daily to distributor servers, where clients use IT-PIR to privately retrieve them.
While PIR hides which messages clients read, the email server stores the communication patterns between email addresses.
Riffle~\cite{kwon2016riffle} uses mixnets to send messages
and IT-PIR to retrieve them. 
Riffle provides k-anonymity; in each round, the adversary learns that
a message originated from 1 of $k$ users.
As discussed in \S\ref{sec:background:goal}, in practice, k-anonymity typically provides weaker privacy than access sequences indistinguishability.

Pung~\cite{angel2016unobservable} (and its refinement~\cite{angel2018})
supports a key-value store based on C-PIR with a security goal of UO-ER,
which is similar to \name's access sequence indistinguishability (see \S\ref{sec:background:threat}).
Pung targets a stronger threat model than \systemname{};
Pung assumes all servers are untrusted.
Hence, Pung only requires one server to provide functionality
compared with $l$ for \name.
Pung's stronger threat model comes at the cost of performance.
Pung uses an implementation of C-PIR for reads called SealPIR~\cite{aguilar2015xpir,angel2018},
which leverages the SEAL homomorphic encryption library~\cite{sealcrypto} 
based on the Fan-Vercauteren fully-homomorphic encryption (FHE) system~\cite{cryptoeprint:2012:144}. 
Thus, Pung incurs orders of magnitude higher computational and network costs,
compared with \systemname{}.

Additionally,
Pung uses an interactive binary search algorithm for retrieval,
requiring $O(log(n))$ round trips between client and server,
compared to the $O(1)$ cuckoo table lookup in \systemname{}.
Thus, to retrieve a 1KB message from a database size of $n=$ 32K messages,
Pung requires $>36MB$ of data.
Pung++ reduces this overhead by leveraging a more efficient
C-PIR protocol, probabilistic batch coding, and \emph{reverse} cuckoo hashing.
The latter places each element into \emph{all} candidate buckets,
and requires each client to have an oracle to tell it which index within a bucket
contains the message it wants to retrieve.
The authors suggest instantiating the oracle by having the client
retrieve a Bloom filter that encodes the index of every message.
Hence, the client must download $O(n)$ data each time it wants
to discover the location of new messages.
Hence,
Pung++ still requires more than $>1MB$ of data for the same workload,
while \systemname{} makes 2 requests and transfers $<12KB$ (10$\times$ less).

\paragraph{Mixnet-based Systems}
Chaum mixnets~\cite{chaumcmix,chaum1981untraceable,gulcu1996mixing,jerichow1998real}
and verifiable cryptographic shuffles~\cite{brickell2006efficient,furukawa2001efficient,neff2003verifiable}
are a way to obfuscate the source of a message.
As with PIR, some mixnet-based systems provide anonymous broadcasting,
rather than private messaging.
For example, Atom \cite{kwon2017atom} implements anonymous broadcast by dividing the servers into multiple groups and then repeatedly shuffling a batch of ciphertexts within each group and forwarding parts of the batch to neighboring groups. 
Atom assumes that some fraction of servers are honest. 
In particular, Atom's security relies on each group having at least one honest server.

Mixnets have been applied to scalable private messaging~\cite{danezis2003mixminion},
but require messages from honest users in every round to form an anonymity set.
When a mixnet is used to access an encrypted database,
unlinkability can be difficult to guarantee when the database is untrusted.
Network-level onion routing systems~\cite{reed1998anonymous,dingledine2004tor,jones2011hiding}
can also be used to access an encrypted database with similar limitations.

Using differential privacy analysis, 
Vuvuzela~\cite{van2015vuvuzela}, Stadium~\cite{tyagi2017stadium}, and Karaoke~\cite{lazar2018karaoke}
formalize the amount of noise that honest shufflers would need to inject
to bound information leakage at the database.
These systems have a weaker security goal than \systemname{} --
which provides indistinguishability even under an active adversary,
but they offer substantially better performance than \systemname{}.
Vuvuzela scales to millions of users with a peak throughput of nearly 4M messages/min
using the same number of servers as \systemname{}.
Stadium~\cite{tyagi2017stadium} achieves Vuvuzela's security goal with even better performance,
but it weakens the threat model to assume that some fraction (the authors suggest 50-75\%) of servers behave honestly.
Karaoke~\cite{lazar2018karaoke} extends Vuvuzela with an efficient noise verification technique
and a security goal of optimistic indistinguishability,
where no information is leaked under a passive adversary, but the system falls back to 
differential privacy under an active attack. 
Similar to Stadium, Karaoke assumes some fraction (the authors suggest 60-80\%) of servers behave honestly.

Like Stadium and Karaoke,
XRD~\cite{kwon2019xrd} is a mix-net system that assumes a fraction of servers behave honestly
and requires similarly large fraction to achieve good performance. 
In contrast to Stadium and Karaoke, XRD offers cryptographic privacy similar to
the guarantees provided by Talek. 

Stadium, Karaoke, and XRD include support for horizontal scaling (linear scaling for Stadium and Karaoke, square root scaling for XRD), 
so that adding more servers improves the system's throughput.
As discussed in \S\ref{sec:eval:throughput},
\name is compatible with instantiating each ``server'' via multiple machines
to provide better throughput.
This provides a subtly different form of scaling, however.
Stadium, Karaoke, and XRD can increase performance by 
increasing the number of participating organizations, 
while \name increases the number of machines held by the participating organizations. 
The difference stems from \name's anytrust threat model.

\paragraph{DC-nets}
DC-net systems~\cite{chaum1988dining,golle2004dining}, 
like Herbivore~\cite{sirer2004eluding} and
Dissent~\cite{corrigan2010dissent,wolinsky2013proactively,wolinsky2012dissent},
are a method for anonymously broadcasting messages to a group using information-theoretic techniques.
For each message, all clients must broadcast random bits to every other client.
DC-nets enable effective broadcast messaging, but they are not optimal for group messaging due to high network costs.

\paragraph{Oblivious RAM (ORAM)}
ORAM~\cite{goldreich1987towards,goldreich1996software,ostrovsky1990efficient}
allows a single trusted client to access untrusted storage
without revealing access patterns, even to a strong adversary who controls the storage.
High network costs of reads, $\Omega(\log n)$,
and constant data reshuffling make ORAM costly for systems with many users sharing data.

 \section{Conclusion}\label{sec:conclusion}
\FloatBarrier{}

We have explored a new point in the design space of private group messaging:
a very strong security guarantee based on access sequence indistinguishability
coupled with an only slightly weakened anytrust threat model.
The result is a new built system, \name, which incorporates
a careful series of design decisions and new optimizations,
including IT-PIR based on blocked cuckoo hashing,
serialized PIR, private notifications, and GPU-based acceleration.
Together, these demonstrate that this design point
supports strong performance for realistic workloads.

\begin{acks}
This work was supported by Google, the Alfred P. Sloan Foundation, and the CONIX Research Center, one of six centers in JUMP, a Semiconductor Research Corporation (SRC) program sponsored by DARPA. This work was also supported by DARPA and NIWC under contract N66001-15-C-4065. The U.S. Government is authorized to reproduce and distribute reprints for Governmental purposes not withstanding any copyright notation thereon. The views, opinions, and/or findings expressed are those of the author(s) and should not be interpreted as representing the official views or policies of the Department of Defense or the U.S. Government.
\end{acks}

\clearpage

\bibliographystyle{ACM-Reference-Format}
\bibliography{paper}

\appendix

\section{Security Proofs}

\subsection{Building Blocks}\label{sec:security:definitions}
We provide formal definitions of the cryptographic building blocks that we use in the security proof of our construction.

We define a Private Information Retrieval scheme as specified by Chor et al.~\cite{chor1995private}, adapted to allow up to $t$ servers to collude (whereas they assume no collusion).
The scheme allows a user, on a desired index $i$ and a random input $r$ of length $l_{rnd}$, to produce $k$ queries of length $l_q$ (one for each server). The servers respond according to the strategies $A_1,\ldots , A_k$ with replies of length $l_a$. The user then reconstructs the desired bit of the database based on these replies. 

\begin{definition}{Private Information Retrieval – One-Round Schemes.}
A $k$-server Private Information Retrieval (PIR) scheme for a database of length $n$ consists of:
\begin{itemize}
\item $k$ query functions, $Q_1,\ldots ,Q_k : [n] \times \{0, 1\}^{l_{rnd}}\rightarrow \{0,1\}^{l_q}$
\item $k$ answer functions, $A_1,\ldots , A_k : \{0, 1\}^n\times \{0,1\}^{l_q} \rightarrow \{0,1\}^{l_a}$
\item a reconstruction function, $R : [n] \times \{0, 1\}^{l_{rnd}} \times (\{0,1\}^{l_a})^k \rightarrow \{0,1\}$
\end{itemize}
These functions should satisfy

\textbf{Correctness:} For every $x \in \{0,1\}^n, i\in [n],$ and $r \in \{0,1\}^{l_{rnd}}$
\begin{gather*}
R(i,r,A_1(x,Q_1(i,r)),\ldots ,A_k(x,Q_k(i,r))) = x_i
\end{gather*}

\textbf{$t$-Privacy:} For every $i,j \in [n], (q_1,\ldots,q_t) \in (\{0,1\}^{l_q})^t$ and $\{s_1,\ldots,s_t\} \subset \{1,\ldots, k\}$
\begin{gather*}
\mathbf{Pr}((Q_{s_1}(i,r),\ldots ,Q_{s_t}(i,r))=(q_1,\ldots ,q_t))=\\
\mathbf{Pr}((Q_{s_1}(j,r),\ldots ,Q_{s_t}(j,r))=(q_1, \ldots ,q_t))
\end{gather*}
where the probabilities are taken over a uniformly chosen $r\in \{0,1\}^{l_{rnd}}$.
\end{definition}
Note that while this definition is for a one-round PIR scheme, it can be extended to multiple rounds where the queries in each round are allowed to depend on answers received in previous rounds.

We use the following standard definitions of a pseudorandom function and the IND-CCA security of the encryption scheme:
\begin{definition}{Pseudorandom function}

A function $F:\{0,1\}^n\times \{0,1\}^{m_1} \rightarrow \{0,1\}^{m_2}$ is called a pseudorandom function (PRF) if it satisfies the following:

\begin{itemize}
\item For all $s\in \{0,1\}^n$ and all $x\in \{0,1\}^{m_1}$, $F(s,x)$ can be computed in polynomial time.

\item Any PPT adversary $\mathcal{A}$ is successful in the following game with probability at most $\frac{1}{2}+negl(n)$, where $negl(\cdot )$ is a negligible function:

\begin{itemize}
\item The challenger chooses $s\leftarrow \{0,1\}^n$ and a bit $b\leftarrow \{0,1\}$ at random.
\item For $i=1,\ldots,q$, where $q$ is polynomial in $n$, $D$ chooses $x_i \in \{0,1\}^{m_1}$ and sends it to the challenger.
\item If $b=0$, the challenger replies with $F(s,x_i)$. Otherwise, if $x_i$ has not been queried before, the challenger picks $y_i \in \{0,1\}^{m_2}$ uniformly at random and sends it to $\mathcal{A}$. If $x_i$ has been queried previously, the challenger sends the same response as the last time $x_i$ was queried.
\item $\mathcal{A}$ outputs $b^*\in \{0,1\}$ and wins if $b^*=b$.
\end{itemize}
\end{itemize}

\end{definition}

\begin{definition}{IND-CCA security}

An encryption scheme $\mathit{(Gen, Enc, Dec)}$ is said to be IND-CCA secure if any PPT adversary $\mathcal{A}$ is successful in the following game with the probability at most $\frac{1}{2}+negl(n)$, where $negl(\cdot )$ is a negligible function:
\begin{itemize}
\item The challenger generates keys $(pk, sk) \leftarrow Gen(1^n)$.
\item $\mathcal{A}$ receives $pk$ as input.
\item $\mathcal{A}$ gets a black-box access to $Dec_{sk}(\cdot )$.
\item $\mathcal{A}$ chooses $x_0, x_1$.
\item The challenger chooses $b\leftarrow \{0,1\}$ at random and gives $\mathcal{A}$ the challenge ciphertext $c=Enc_{pk}(x_b)$.
\item $\mathcal{A}$'s access to $Dec_{sk}(\cdot )$ is now restricted - $\mathcal{A}$ is not allowed to ask for the decryption of $c$.
\item $\mathcal{A}$ outputs $b^*\in \{0,1\}$ and wins if $b=b^*$.
\end{itemize}
 
\end{definition}

Note that authenticated encryption implies single-message IND-CCA security,
and single-message IND-CCA security implies multi-message IND-CCA security.

Finally, we are using the random oracle model where all parties have access to a random oracle, which is defined as follows:
\begin{definition}{Random oracle}

Given a security parameter $n$ and a length function $l_{out}(\cdot)$, a random oracle $R$ is a map from $\{0,1\}^*$ to $\{0,1\}^{l_{out}(n)}$.
\end{definition}

\subsection{Proof of Access Sequence Indistinguishability}\label{sec:security:proof}
We provide a security proof for \systemname's
protocol by reduction to the cryptographic assumptions provided in Appendix~\ref{sec:security:definitions}. Definitions of $\mathit{RealWrite}$ and $\mathit{FakeWrite}$ are provided in~\S\ref{sec:design:cuckoo}. $\mathit{RealRead}$, $\mathit{FakeRead}$ are defined in~\S\ref{sec:design:pir}. Finally, interest vectors (addressed in {\bf Game 4}) are defined in~\S\ref{sec:design:notifications}
.

We consider a series of games adapted from the game in~\S\ref{sec:secgame},
each defined from the previous one by idealizing some part of the protocol.
For game $i$, we write $p_i$ for the maximum advantage, $|Pr(b=b') - 1/2|$,
that $\mathcal{A}$ holds in the security game.
At each step, we bound the adversary's advantage between two successive games.
Technically, each of the following games consists of a series of hybrid games,
where we change each of the $m$ clients one by one.

\vspace{0.05in}
\noindent {\bf Game 0:}
This is the original game defined in~\S\ref{sec:secgame} with an adversary $\mathcal{A}$
that chooses $m$ challenger clients,
and submits sequences with $\alpha_0$ calls to $\mathit{RealRead}$, and
$\alpha_1$ calls to $\mathit{RealWrite}$,
using the protocol defined in the paper.
All $\mathit{RealRead}$s and $\mathit{FakeRead}$s from the client trigger $\mathit{ProcessRead}$ on the servers.
All $\mathit{RealWrite}$s and $\mathit{FakeWrite}$s from the client trigger $\mathit{ProcessWrite}$ on the servers.
Thus, the adversary has control over the messages sent by the clients over the network.
We intend to show that the adversary's advantage in this game is negligible.

\vspace{0.05in}
\noindent {\bf Game 1: (PIR $\mathit{Read}$)}
This game is as above, except that in each $\mathit{RealRead}(\tau , seqNo)$, we replace the real read location (specifically the bucket $PRF(\tau .k, seqNo)$) by the index of an arbitrary bucket, as specified in $\mathit{FakeRead}$.
Let $\epsilon^{PIR}(\lambda_0,n)$ bound the advantage of an adversary
breaking the PIR assumption in $n$ calls to PIR read (via $\mathit{RealRead}$) with security parameter $\lambda_0$.

Given an adversary $\mathcal{G}$ that distinguishes between {\bf Game 0} and {\bf Game 1} and makes $n$ calls to $\mathit{RealRead}$, we can construct an adversary $\mathcal{P}$ on the security of the $n$-round PIR scheme. $\mathcal{P}$ starts by sending the set of indices of the servers, controlled by $\mathcal{G}$ to its own challenger $\mathcal{C}_P$. Then, $\mathcal{P}$ behaves like a challenger to $\mathcal{G}$, following the description of the access sequence indistinguishability game, except for the case where a read request must be issued and the read queue is not empty (thus, a $\mathit{RealRead}(\tau, seqNo)$ is executed). In this case, $\mathcal{P}$ computes $i:=PRF(\tau.k, seqNo)$, where $k$ is equal to $k_{s1}$ or $k_{s2}$, depending on whether the first or the second cuckoo location must be read; $\mathcal{P}$ also chooses $j$ at random in $[0..b-1]$. Then, $\mathcal{P}$ forwards $i$ and $j$ to $\mathcal{C}_P$. Upon receiving the $l-1$ PIR request queries from its challenger in response, $\mathcal{P}$ forwards these to $\mathcal{G}$. Upon $\mathcal{G}$ ending the game, $\mathcal{P}$ responds to its own challenger $\mathcal{C}_P$ with $i$, if $\mathcal{G}$'s response was {\bf Game 0}, and $j$, if $\mathcal{G}$'s response was {\bf Game 1}. Note that $i$ corresponds to the real location specified by the $\mathit{RealRead}$ request, whereas $j$ is sampled at random, as defined by $FakeRead$. Thus, if $\mathcal{C}_P$ chooses index $i$, then the game $\mathcal{G}$ is in is exactly {\bf Game 0}, and if $\mathcal{C}_P$ chooses index $j$, then the game $\mathcal{G}$ is in is exactly {\bf Game 1}. Since the adversary makes $\alpha_0$ calls to $RealRead$, we get:
\begin{center}
  $p_0 \leq p_1 + m \cdot \epsilon^{PIR}(\lambda_0, \alpha_0)$
\end{center}

\vspace{0.05in}
\noindent {\bf Game 2: (IND-CCA with $\mathit{Write}$)}
This game is as above, except that $\mathit{RealWrite}$ is modified to encrypt a dummy message instead of $seqNo | M$. 

Let $\epsilon^{AE}_{IND-CCA}(\lambda_1, n)$ be the advantage of an adversary who breaks the IND-CCA assumption and performs $n$ calls to $\mathit{RealWrite}$ with security parameter $\lambda_1$.

Given an adversary $\mathcal{G}$ that distinguishes between {\bf Game 1} and {\bf Game 2} and makes $n$ calls to $\mathit{RealWrite}$, we can construct an adversary $\mathcal{P}$ on the $n$-message IND-CCA security of the used encryption scheme. $\mathcal{P}$ behaves like a challenger to $\mathcal{G}$, following the description of the access sequence indistinguishability game, except for the case where a write request must be issued and the write queue is not empty (thus, a $\mathit{RealWrite}(\tau, seqNo, M)$ is executed). In this case, $\mathcal{P}$ generates a random message $M^*$ with the same length as $seqNo | M$. Then, $\mathcal{P}$ forwards $seqNo|M$ and $M^*$ to $\mathcal{C}_P$. Upon receiving the challenge ciphertext $c$ in response, $\mathcal{P}$ forwards $\beta_1|\beta_2|c$ to $\mathcal{G}$, where $\beta_1$ and $\beta_2$ are the numbers of the buckets as specified by the $\mathtt{Write}$ request. Upon $\mathcal{G}$ ending the game, $\mathcal{P}$ responds to its own challenger $\mathcal{C}_P$ with $0$ (meaning the challenge was an encryption of $seqNo|M$), if $\mathcal{G}$'s response was {\bf Game 1}, and $1$, if $\mathcal{G}$'s response was {\bf Game 2} (meaning the challenge was an encryption of $M^*$). Note that if $\mathcal{C}_P$ chooses $seqNo|M$, then the game $\mathcal{G}$ is in is exactly {\bf Game 1}, and if $\mathcal{C}_P$ chooses $M^*$, then the game $\mathcal{G}$ is in is exactly {\bf Game 2}. Since the adversary makes $\alpha_1$ calls to $RealWrite$, we can apply the IND-CCA definition and get:

\begin{center}
    $p_1 \leq p_2 + m \cdot \epsilon^{AE}_{IND-CCA}(\lambda_1, \alpha_1)$
\end{center}

\vspace{0.05in}
\noindent {\bf Game 3: (PRF with $\mathit{Write}$)}
This game is as above, except we replace the PRF used to generate
the bucket locations of $\mathit{RealWrite}$s with a truly random function, such that the client submits a \\
$\mathit{FakeWrite}$.
Let $\epsilon_{distinguish}^{PRF}(\lambda_2, n)$
bound the advantage of an adversary breaking the PRF assumption after $n$
calls to the PRF with a security parameter $\lambda_2$. Technically, this game consists of a series of up to $2n$ games: two for each log handle used in a write request. In the following, we describe the game for replacing the PRF used in the generation of $\beta_1$.

Given an adversary $\mathcal{G}$ that distinguishes between {\bf Game 2} and {\bf Game 3} and makes $n$ calls to $\mathit{RealWrite}$, we can construct an adversary $\mathcal{P}$ on the security of the used PRF that is called up to $n$ times. $\mathcal{P}$ behaves like a challenger to $\mathcal{G}$, following the description of the access sequence indistinguishability game, except for the case where a write request must be issued and the write queue is not empty (thus, a $\mathit{RealWrite}(\tau, seqNo, M)$ is executed). In this case, $\mathcal{P}$ forwards $seqNo$ to $\mathcal{C}_P$. Upon receiving the challenge $\beta$ in response, $\mathcal{P}$ uses it instead of $\beta_1$ in the generated $\mathtt{Write}$ request and forwards the request to $\mathcal{G}$ (note that exactly the same procedure can be done for $\beta_2$). Upon $\mathcal{G}$ ending the game, $\mathcal{P}$ responds to its own challenger $\mathcal{C}_P$ with $PRF$, if $\mathcal{G}$'s response was {\bf Game 2}, and $\mathit{Random\: function}$, if $\mathcal{G}$'s response was {\bf Game 3}. Note that if $\mathcal{C}_P$ chooses to use the PRF on the provided input, then the game $\mathcal{G}$ is in is exactly {\bf Game 2}, and if $\mathcal{C}_P$ chooses to use a random function, then the game $\mathcal{G}$ is in is exactly {\bf Game 3}. Since the adversary makes $\alpha_1$ calls to $RealWrite$ and each of these calls contains two calls to the PRF, we conclude:

\begin{center}
  $p_2 \leq p_3 + 2 \cdot m \cdot \epsilon_{distinguish}^{PRF}(\lambda_2, \alpha_1)$
\end{center}

\vspace{0.05in}
\noindent {\bf Game 4: (Hash functions in interest vectors)}
This game is as above, except we replace the $h$ cryptographic hash functions used in the
Bloom filter of the interest vector with queries to a random oracle. Technically, this game consists of a series of hybrid games, where we change each of the $h$ hash function one by one.
Let $\epsilon^{hash}(\lambda_3,n)$
bound the advantage of an adversary breaking the random oracle assumption in $n$ calls with a security parameter, $\lambda_3$.

Given an adversary $\mathcal{G}$ that distinguishes between {\bf Game 3} and {\bf Game 4} and makes $n$ calls to $\mathit{RealWrite}$, we can construct an adversary $\mathcal{P}$ on the random oracle assumption that is called $n$ times. First, $\mathcal{P}$ chooses $n$ log IDs at random. $\mathcal{P}$ behaves like a challenger to $\mathcal{G}$, following the description of the access sequence indistinguishability game, except for the case where a write request must be issued and the write queue is not empty (thus, a $\mathit{RealWrite}(\tau, seqNo, M)$ is executed). In this case, $\mathcal{P}$ selects one of the $n$ log IDs chosen in the beginning (using the same log id if the log handle has been used in a $\mathit{RealWrite}$ before) and forwards it to $\mathcal{C}_P$. The $\mathtt{Write}$ is then generated by $\mathcal{C}_P$ as specified in the previous game, except that the generation of the interest vector submitted with this request is changed. The interest vector is generated as specified in the previous game, except for the hash function in question. The position in the interest vector defined by this hash function is defined by the challenge received from $\mathcal{C}_P$. Upon receiving this challenge $pos$, $\mathcal{P}$ puts a 1 in the position $pos$ of the interest vector of the generated $\mathtt{Write}$ request and forwards the request to $\mathcal{G}$. Upon $\mathcal{G}$ ending the game, $\mathcal{P}$ responds to its own challenger $\mathcal{C}_P$ with $\mathit{Hash\: function}$, if $\mathcal{G}$'s response was {\bf Game 3}, and it responds with $\mathit{Random\: oracle}$, if $\mathcal{G}$'s response was {\bf Game 4}. Note that if $\mathcal{C}_P$ chooses to use the hash function on the input provided, then the game $\mathcal{G}$ is in is exactly {\bf Game 3}, and if $\mathcal{C}_P$ chooses to use a random oracle, then the game $\mathcal{G}$ is in is exactly {\bf Game 4}. Since the adversary makes $\alpha_1$ calls to $\mathit{RealWrite}$, and each of these calls contains one call to each of the $h$ cryptographic hash functions, we conclude:
\begin{center}
  $p_3 \leq p_4 + m \cdot h \cdot \epsilon^{hash}(\lambda_3, \alpha_1)$
\end{center}

From this final game,
all of the parameters in any network request have been replaced with random values.
Because Game 4 involves all clients issuing periodic requests with
random parameters, by definition the
adversary's advantage, $p_4$, must be negligible.

\vspace{0.05in}
\noindent {\bf Privacy:}
Collecting the probabilities from all games yields:
\begin{align*}
  p_0 \leq & m \cdot \epsilon^{PIR}(\lambda_0, \alpha_0) + \\
           & m \cdot \epsilon^{AE}_{IND-CPA}(\lambda_1, \alpha_1) + \\
           & 2 \cdot m \cdot \epsilon_{distinguish}^{PRF}(\lambda_2, \alpha_1) + \\
           & m \cdot h \cdot \epsilon^{hash}(\lambda_3, \alpha_1)
\end{align*}
$p_0$ becomes negligible for large security parameters $\lambda_0$, $\lambda_1$, $\lambda_2$, and $\lambda_3$.

\paragraph{Proof of Security for Serialized PIR}\label{sec:security:serializedpirproof}
We provide intuition for the security of the serialized PIR introduced in \S\ref{sec:serialized} by reduction to the security of the underlying PIR system discussed in \S\ref{sec:background:pir}, and the cryptographic assumptions of the cryptographic primitives used. First, because we use authenticated encryption, sending encryptions of the seed $p_i$ and PIR request values $q_i$ for each server $i$ through the lead server is as secure as sending values $p_i$, $q_i$ to server $i$ directly through authenticated secure communication channels. Next, because of the security of the PRNG scheme, sending values $p_i$ and computing $P_i=PRNG(p_i)$ on the server side is comparable to the security of the client generating random bit vectors $P_i$ and sending those to the servers instead. Finally, because at this point vectors $P_i$ are randomly generated and can be viewed as a one-time pad, computing responses $R_i$ and sending values $R_i \oplus P_i$ to the lead server is as secure as computing and sending responses $R_i$ to the client directly. At this point, the modified protocol corresponds exactly to the original version discussed in~\S\ref{sec:background:pir}.

\end{document}